\documentclass{aa}

\usepackage{graphicx}

\usepackage{txfonts}
\usepackage{lscape}

\begin{document}

   \title{The Moon at thermal infrared wavelengths:  A benchmark for asteroid thermal models}
   \author{
           T.\ G.\ M\"{u}ller \inst{\ref{inst1}} \and
           M.\ Burgdorf \inst{\ref{inst2}} \and
           V.\ Al{\'i}-Lagoa \inst{\ref{inst1}} \and
           S.\ A.\ Buehler \inst{\ref{inst2}} \and
           M.\ Prange \inst{\ref{inst2},\ref{inst3}}
          }
   \institute{   {Max-Planck-Institut f\"{u}r extraterrestrische Physik,
                 Giessenbachstra{\ss}e, Postfach 1312, 85741 Garching, Germany
                 \email{tmueller@mpe.mpg.de}\label{inst1}
                 }
                 \and
                 {Meteorologisches Institut, Centrum f\"ur Erdsystem- und Nachhaltigkeitsforschung
                 (CEN), Universit\"at Hamburg, Bundesstrasse 55, 20146 Hamburg, Germany
                 \label{inst2}
                 }
                 \and
                 {International Max Planck Research School on Earth System Modelling (IMPRS-ESM),
                 Bundesstra{\ss}e 53, 20146 Hamburg, Germany\label{inst3}
                 }
          }
   \date{Received ; accepted }
\abstract{Thermal-infrared measurements of asteroids, satellites, and distant minor bodies are
          crucial for deriving the objects' sizes, albedos, and in some cases, also the thermophysical
          properties of the surface material. Depending on the available measurements and auxiliary
          data, such as visual light curves, spin and shape information, or direct size measurements
          from occultations or high-resolution imaging techniques, a range of simple to complex thermal models are applied to achieve specific science goals. However, testing
          these models is often a difficult process and the uncertainties of the derived parameters are not
          easy to estimate. Here, we make an attempt to verify a widely accepted thermophysical model (TPM)
          against unique thermal infrared (IR), full-disk, and well-calibrated measurements of the Moon.
          The data were obtained by the High-resolution InfraRed Sounder (HIRS) instruments on board
          a fleet of Earth weather satellites that  serendipitously scan the surface of the Moon.
          We found 22 Moon intrusions, taken in 19 channels between 3.75\,$\mu$m
          and 15.0\,$\mu$m, and over a wide phase angle range from -73.1$^{\circ}$ (waxing Moon)
          to +73.8$^{\circ}$ (waning Moon).
          These measurements include the entire Moon in a single pixel, seen almost simultaneously
          in all bands. The HIRS filters are narrow and outside the wavelength regime of the
          Christiansen feature.
          The similarity between these Moon data and typical asteroid spectral-IR energy
          distributions allows us to benchmark the TPM concepts and to point out problematic aspects.
          The TPM predictions match the HIRS measurements within 5\% (10\% at the shortest wavelengths
          below 5\,$\mu$m) when using the Moon's known properties (size, shape, spin, albedo, thermal
          inertia, roughness) in combination with a newly established wavelength-dependent hemispherical
          emissivity. In the 5-7.5\,$\mu$m and in the 9.5 to 11\,$\mu$m ranges, the global
          emissivity model deviates considerably from the known lunar sample spectra.
          Our findings will influence radiometric studies of near-Earth and main-belt asteroids
          in cases where only short-wavelength data (from e.g., NEOWISE, the warm Spitzer mission,
          or ground-based M-band measurements) are available.
          The new, full-disk IR Moon model will also be used for the calibration of
          IR instrumentation on interplanetary missions (e.g., for Hayabusa-2) and
          weather satellites.
         }
\authorrunning{M\"uller et al.}
\titlerunning{Benchmarking the asteroid TPM against the Moon}
\maketitle

\section{Introduction}
\label{sec:intro}

Thermophysical modeling techniques are widely used to derive radiometric properties of
asteroids \citep[e.g.,][]{Delbo2015} and more distant bodies
\citep[e.g.,][]{Mueller2020}. Most of the radiometric studies are based
on infrared (IR) measurements taken close to the objects' thermal emission peak:
IRAS  \citep{Tedesco2002} at 12, 25, 60, and 100\,$\mu$m,
MSX \citep{Tedesco2002b} at 4.29, 4.35, 8.28, 12.13, 14.65, and 21.3\,$\mu$m,
AKARI-IRC \citep{Usui2011,Usui2013} at 9 and 18\,$\mu$m, or
WISE/NEOWISE \citep{Mainzer2011b,Mainzer2016} at 11 and 22\,$\mu$m.
For more distant, colder Centaurs or trans-Neptunian objects, key studies used
data from Spitzer-MIPS at 24 and 70\,$\mu$m \citep{Stansberry2008}, or from Herschel-PACS
at 70, 100, and 160\,$\mu$m \citep{Mueller2020}.

A comparison between radiometric sizes and true sizes (from occultations, direct
imaging, or in-situ studies) shows an excellent agreement
\citep[e.g.,][and references therein]{Harris2002,Usui2014,Mainzer2015,Ortiz2020}.
Deviating radiometric sizes are typically attributed to poor spin-shape information, 
low-quality or single-epoch thermal observations, or a combination of these
aspects. 

Recently, more and more IR data of asteroids at shorter wavelengths (far away from 
the thermal emission peak) have become available:
NEOWISE W1 and W2 bands at 3.4 and 4.6\,$\mu$m after the end of the cryogenic
mission phase \citep{Mainzer2014a}, 
Spitzer-IRAC at 3.6 and 4.5\,$\mu$m warm mission (2009-2020) \citep{Mahoney2010},
or ground-based observations up to about 5\,$\mu$m, for example, using SpeX at NASA's
Infrared Telescope Facility \citep{Moskovitz2017}. These short-wavelengths measurements are
often focused on near-Earth objects and smaller main-belt asteroids with the goal to
obtain the fundamental object properties. However, the radiometric model techniques
are not well tested below 10\,$\mu$m.

In general, there are no benchmark targets with consistently high-quality disk-integrated
thermal measurements over a wide range of wavelengths and phase angles. 
Either the few objects with accurate or in-situ physical properties are too bright for
the IR instruments or the disk-integrated thermal IR data available for those targets
are insufficient to independently tackle TPM validation over a wide range of observational
configurations and its large model parameter space (e.g.,\ the Hayabusa mission target
25143~Itokawa). 
Our Moon has been an important ground-truth reference in terms of reproducing instantaneous
surface temperatures. However, a systematic comparison with asteroid thermophysical studies
based on non-resolved data has been more difficult due to instrument saturation problems
or the predominance of disk-resolved data that cover only part of the Moon. 

For example, the infrared radiometer Diviner \citep{Paige2010} on board NASA's Lunar
Reconnaissance Orbiter (LRO) obtained large amounts of IR measurements and was able to create
temperature maps of the Moon's surface. However, a direct comparison between Diviner
products and TPM predictions would be restricted to very limited phase angle and
wavelength coverage\footnote{In the thermal IR, Diviner has three narrow-band mineralogy
channels at 7.8, 8.2, and 8.6\,$\mu$m and four broad-band thermal channels
covering the 12.5 to 200\,$\mu$m range.}.
Furthermore, with the Diviner data it would not be possible to test our TPM over
typical asteroid phase angles and in the short-wavelength thermal regime below
7.5 $\mu$m.

Here, we present a unique data set of full-disk Moon measurements in the wavelength
range between 3.75 and 15\,$\mu$m, obtained by a fleet of weather satellites, each carrying
a High Resolution Infrared Radiation Sounder (HIRS) instrument (Section~\ref{sec:hirs}). We
interpret these measurements with a thermophysical model (TPM), which is widely
used in the context of minor body studies (Section~\ref{sec:tpm}).
The comparison between our TPM predictions and the HIRS measurements are shown in
Section~\ref{sec:comp}. The discussion, along with several potential applications
of our Moon model, is presented in Section~\ref{sec:dis}, followed by our
conclusions in Section~\ref{sec:con}.


\section{Disk-integrated thermal-IR measurements of the Moon}
\label{sec:hirs}

We carried out our investigations with data from HIRS\footnote{High-resolution InfraRed Sounder}
instruments, which are scanning radiometers that perform operational atmospheric sounding.
They are part of the TOVS sounding instrument suite
(TIROS\footnote{Television InfraRed Observation Satellite} Operational Vertical Sounder)
and have evolved up to HIRS/4.
In the course of this evolution, the diameter of its field of view (FOV)
decreased from 1.4$^{\circ}$ to 0.7$^{\circ}$\footnote{This decrease reduces the
instantaneous FOV size from 20 to 10\,km at the sub-satellite point. The scanning technique,
however, remained the same with steps of 26\,km cross-track and 42\,km along-track.}.
HIRS has 19 infrared channels (3.8-15\,${\mu}$m)
and one visible channel, which is not relevant for our study.
The long-wavelength (LW, ch01-12) and short-wavelength (SW, ch13-ch19) channels
have different optical paths. There is a displacement of the channels perpendicular
to the scan direction. This displacement is different for LW and SW, but in either
case, it is proportional to wavelength \citep{Burgdorf2020}. All channels, therefore, have
slightly different view directions, which might sometimes result in the Moon
not being fully included in the FoV of all LW or SW channels above or below a certain
wavelength.

\begin{figure}[h!tb]
\resizebox{\hsize}{!}{\includegraphics{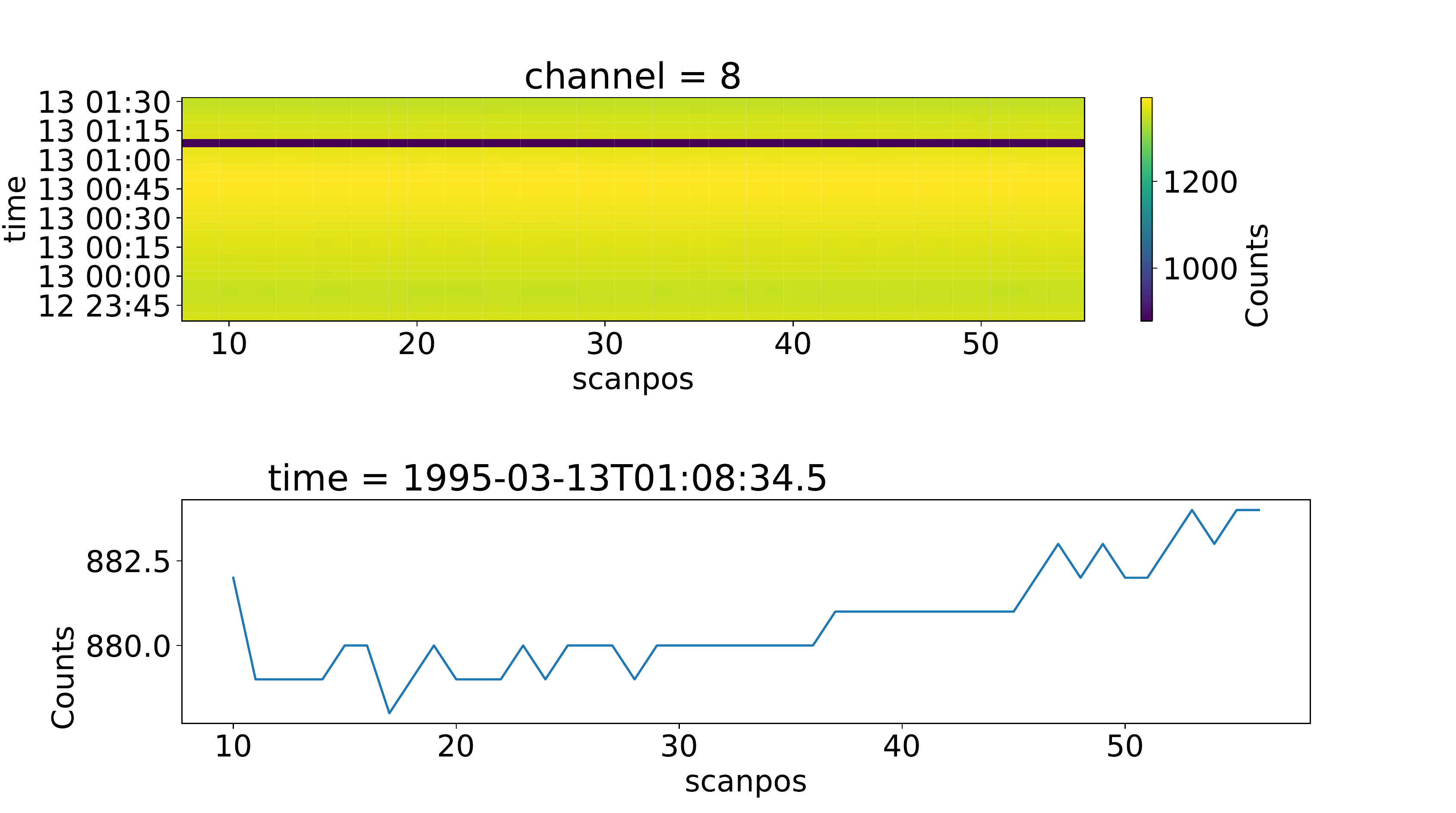}}
\caption{Signal from channel 8 of HIRS/2 on NOAA-14 while viewing deep
         space during one complete orbit on March 12-13, 1995 (top). Here,
         "scanpos" is equivalent to "sample"; all 56 samples of one calibration sequence
         are taken within 6.4\,sec. As HIRS gives fewer counts for stronger flux, any
         intrusion of the Moon in the field of view is easily recognizable by the
         exceptionally low number of counts. The number of counts from the scan
         with the Moon in the FOV is plotted in the bottom panel, where it is
         possible to see that the Moon was fully included in the FOV until
         scanpos 33. Then it began to slowly move out. At scanpos 10, the scanning
         mirror had not yet reached its final position and, therefore, the Moon
         was not fully included in the FOV.
         \label{fig:fig_hirs_signal}}
\end{figure}

The two point IR calibration of HIRS is provided by programmed views of two radiometric targets: the warm target, mounted on the
instrument baseplate, and a view of deep space (DSV). Each view takes 6.4\,sec, including the
time required to bring the scanning mirror into position. Data from these views provide
sensitivity calibrations for each channel at 256 second intervals. The deep-space view points
at a direction close to the orbital axis of the satellite, that is, near the celestial equator.
This means that occasionally the Moon appears in the DSV at the time of the calibration.
Its presence corrupts the signal, which normally corresponds to zero flux.
In most cases, the signal from the Moon changes over the 6.4\,sec while the instrument is
viewing space because the Moon is moving in or out of the FOV, but when its signal is
constant, the Moon must be fully included in the FOV and none of its flux is lost.
Such events were tracked down with the help of
STAR ICVS\footnote{\url{https://www.star.nesdis.noaa.gov/icvs/index.php}}
(Integrated Calibration / Validation System Long-Term Monitoring) and identified in
the level 1b data from eight different satellites in the NOAA CLASS (Comprehensive
Large Array-data Stewardship System) archive (see Figure~\ref{fig:fig_hirs_signal}).
The HIRS Level 1b datasets provide raw instrument counts. Once we found a suitable
Moon intrusion in the DSV, we used the observations of the warm target and of the DSV
before and after for the IR calibration of HIRS.

The noise of the measurements was calculated directly from the distribution of
the counts in each channel. The bottom panel of Fig.~\ref{fig:fig_hirs_signal} demonstrates
the impact of digitization on the PDF\footnote{Probability Density Function} shape.
On top of the pure random variations
of the counts, there is also the structured effect of short-term temperature changes
of a baffle contributing to the counts from the receiver \citep{Labrot2019}.
Developing a self-emission model for HIRS in order to correct for these temperature
changes, however, is hardly worthwhile in light of the small number of Moon
intrusions we identified. The additional uncertainty caused by this effect
was instead estimated from the difference of the flux values measured at very
similar phase angles of the Moon and from the scatter of the counts in the DSV
calibration measurements when the Moon was absent. While the size of the independent
noise is usually negligible, the temperature changes lead to an uncertainty of some
2\,K in the measured brightness temperatures. 

There are several common effects that limit the accuracy of the measurements.
Band corrections for each IR channel are used to modify the temperature of the
warm target as measured with the platinum resistance thermometers to give an
effective temperature. The effective temperature is for use in the Planck function
so that the correct radiance is obtained with the central wavelengths
\citep[see][and the corresponding software
documentation\footnote{\url{https://www.nwpsaf.eu/site/software/aapp/documentation/}}]{Labrot2019}.

The spectral response functions for each HIRS channel had to be optimized
to generate a more consistent set of observations \citep{Chen2013}.
Shift values were therefore used as provided by
E.\ Borbas\footnote{\url{https://www.nwpsaf.eu/site/software/rttov/download/coefficients/spectral-response-functions/}}.

We assumed that radiance can be related to counts through
a linear equation. This assumption can be made because the
non-linearity term in the HIRS radiometric calibration was found
to be much smaller in flight than the pre-launch value
\citep{Chen2012},
and the ATOVS\footnote{Advanced TIROS Operational Vertical Sounder}
and AVHRR\footnote{Advanced Very High Resolution Radiometer}
Pre-processing Package (AAPP) does not
perform a non-linearity correction either. 

The diameter of the FOV was assumed to be 1.4$^{\circ}$ for HIRS/2,
1.3$^{\circ}$ for HIRS/3, and 0.7$^{\circ}$ for HIRS/4 according to
OSCAR (Observing Systems Capability Analysis and Review Tool)\footnote{\url{http://www.wmo-sat.info/oscar/}}.
These numbers were verified by comparing the flux values measured
with different satellites at very similar phase angles of the Moon. The
uncertainty of the diameter of the FOV causes an uncertainty of the
measured fluxes of at most a few percent. All channels 1-12 or 13-19 of a given
satellite are affected in the same way.

In summary, our processing of the raw data from the Moon intrusions used
the same method as AAPP, namely:\ using a two-point calibration interpolating between
the signals from the warm (between 283.72\,K and 285.97\,K, depending on
instrument setting) and the cold target (space background), but without
taking the changing flux contributions from the baffle into account
\citep[see also][]{Burgdorf2020}.
Table~\ref{tbl:hirs_events} lists the events when the Moon was completely in the FOV
(at least for a subset of channels). Table~\ref{tbl:moon_geom} summarizes the observing
geometry as seen from the satellites, together with apparent Moon properties.

  \begin{table*}[h!tb]
    \begin{center}
    \caption{Overview of HIRS measurements where the Moon was completely in the FOV. The satellite
             longitude (positive values: east, negative values: west), latitude (positive values: north,
             negative values: south), and altitude values (in kilometers above the Earth's surface)
             are needed to calculate the correct viewing geometry for the Moon
             (distance, phase angle, and apparent diameter).
             \label{tbl:hirs_events}}
    \begin{tabular}{llrrrrrr}
      \hline
      \hline
      \noalign{\smallskip}
                &        & Julian     &  Date       & Time   & longitude    & latitude     & altitude \\
      Satellite & Instr. & Date       &  YYYY-MM-DD & HH:MM  & [$^{\circ}$] & [$^{\circ}$] & [km] \\
      \noalign{\smallskip}
      \hline
      \noalign{\smallskip}
      NOAA-11 & HIRS2 & 2449317.23542 &  1993-11-25 &  17:39 & +157.5 & -11.95 & 852 \\
      NOAA-11 & HIRS2 & 2448668.32917 &  1992-02-15 &  19:54 &  -66.4 & -43.19 & 845 \\
      NOAA-11 & HIRS2 & 2448222.71111 &  1990-11-27 &  05:04 &  -38.2 & +23.43 & 850 \\
      NOAA-11 & HIRS2 & 2447602.64653 &  1989-03-17 &  03:31 & +166.6 & -56.45 & 887 \\
      NOAA-14 & HIRS2 & 2451648.95139 &  2000-04-14 &  10:50 & -105.3 & -14.38 & 856 \\
      NOAA-14 & HIRS2 & 2450791.41111 &  1997-12-08 &  21:52 &  +73.3 & +24.68 & 849 \\
      NOAA-14 & HIRS2 & 2450614.71181 &  1997-06-15 &  05:05 &  -31.8 & +48.97 & 867 \\
      NOAA-14 & HIRS2 & 2450379.52083 &  1996-10-23 &  00:30 & -167.2 & +47.91 & 867 \\
      NOAA-14 & HIRS2 & 2450232.37361 &  1996-05-28 &  20:58 & -120.1 & +56.81 & 866 \\
      NOAA-14 & HIRS2 & 2450054.80069 &  1995-12-03 &  07:13 &  +99.8 &  -8.31 & 858 \\
      NOAA-15$^{a}$
              & HIRS3 & 2458831.90792 &  2019-12-14 &  09:47 & +151.3 & -53.35 & 832 \\
      NOAA-15$^{b}$
              & HIRS3 & 2458476.98889 &  2018-12-24 &  11:44 & +114.1 & -43.93 & 822 \\
      NOAA-17$^{c}$
              & HIRS3 & 2452663.55486 &  2003-01-24 &  01:19 &  -56.0 & +44.19 & 813 \\
      NOAA-17$^{d}$
              & HIRS3 & 2452543.79236 &  2002-09-26 &  07:01 &  +39.8 & -33.81 & 836 \\
      NOAA-18 & HIRS4 & 2458449.24931 &  2018-11-26 &  17:59 &  +41.7 & -41.40 & 865 \\
      NOAA-18 & HIRS4 & 2458210.89375 &  2018-04-02 &  09:27 &  +53.6 & +80.63 & 868 \\
      NOAA-18 & HIRS4 & 2455634.93889 &  2011-03-14 &  10:32 & -134.8 & -46.89 & 882 \\
      NOAA-19 & HIRS4 & 2458090.03403 &  2017-12-02 &  12:49 &  +44.5 & -16.96 & 869 \\
      NOAA-19$^{e}$
              & HIRS4 & 2455990.71324 &  2012-03-04 &  05:07 & +132.2 & -34.69 & 858 \\
      MetOp-A & HIRS4 & 2455882.48750 &  2011-11-16 &  23:42 &  -31.7 & -11.60 & 825 \\
      MetOp-B & HIRS4 & 2458685.83125 &  2019-07-21 &  07:57 &  +94.2 & +80.74 & 833 \\
      MetOp-B & HIRS4 & 2457060.71667 &  2015-02-07 &  05:12 &  +70.6 & +33.23 & 824 \\
      \noalign{\smallskip}
      \hline
    \end{tabular}
            \\
    \footnotesize{$^{a}$ Moon was only fully included in the FOV in ch18 and ch19.
                  $^{b}$ Possibly touching edge of ch08-ch12 FOV.
                  $^{c}$ Possibly touching edge of ch08-ch19 FOV.
                  $^{d}$ Possibly touching edge of ch01-ch19 FOV.
                  $^{e}$ Moon was only in ch18 and ch19 in the FOV.}
    \end{center}
  \end{table*}

  \begin{table*}[h!tb]
    \begin{center}
    \caption{Overview of the Moon geometry during the HIRS measurements.
             APmag is the apparent visual magnitude of the Moon; $\Delta$ is
             the satellite-Moon distance in kilometer; $\alpha$ the phase angle
             (Sun-Moon-Satellite): positive values if the Moon is leading the Sun
             and negative values when the Moon is trailing the Sun; the diameter of
             the Moon's full disk is given via the equatorial angular width (in arcsec
             and steradian), Ob-lon and Ob-lat give 
             the apparent sub-observer longitude and latitude values. The sub-Solar
             longitude and latitude is not listed, but the sub-Solar latitude varies
             only between -1.6 and +1.5$^{\circ}$.
             \label{tbl:moon_geom}}
    \begin{tabular}{crrrrrr}
      \hline
      \hline
      \noalign{\smallskip}
      Date \&          & APmag  & $\Delta$ & $\alpha$     & diameter                     & Ob-lon       & Ob-lat \\
      Time             & [mag]  & [km]     & [$^{\circ}$] & [$^{\prime \prime}$] \& [sr] & [$^{\circ}$] & [$^{\circ}$] \\
      \noalign{\smallskip}
      \hline
      \noalign{\smallskip}
      1993-11-25 17:39 & -11.57 & 4.06832E+05 & -40.1 & 1761.737  5.72946e-05 & 359.1 &  -4.8 \\
      1992-02-15 19:54 & -11.96 & 3.62777E+05 & -34.6 & 1975.683  7.20553e-05 & 357.5 &  +0.6 \\
      1990-11-27 05:04 & -10.83 & 3.80571E+05 & -70.8 & 1883.305  6.54746e-05 & 351.2 &  -5.4 \\
      1989-03-17 03:31 & -11.11 & 3.98021E+05 & -57.5 & 1800.737  5.98594e-05 &   6.8 &  -4.9 \\
      2000-04-14 10:50 & -11.31 & 3.79167E+05 & -53.7 & 1890.279  6.59604e-05 &   2.9 &  -3.3 \\
      1997-12-08 21:52 & -10.93 & 3.71395E+05 & -69.2 & 1929.837  6.87500e-05 & 358.7 &  +2.7 \\
      1997-06-15 05:05 & -10.74 & 4.00922E+05 & -68.0 & 1787.708  5.89963e-05 & 355.4 &  -1.5 \\
      1996-10-23 00:30 & -11.58 & 3.70828E+05 & -46.3 & 1932.785  6.89602e-05 &   2.1 &  -1.8 \\
      1996-05-28 20:58 & -11.32 & 3.86040E+05 & -51.1 & 1856.626  6.36327e-05 & 354.0 &  +0.6 \\
      1995-12-03 07:13 & -11.60 & 3.97263E+05 & -40.8 & 1804.172  6.00880e-05 &   6.8 &  +0.7 \\
      2019-12-14 09:47 & -12.06 & 3.77946E+05 &  28.0 & 1896.384  6.63883e-05 & 356.5 &  -2.2 \\
      2018-12-24 11:44 & -12.23 & 3.63363E+05 &  24.8 & 1972.498  7.18232e-05 &   0.4 &  -0.8 \\
      2003-01-24 01:19 & -10.79 & 3.71946E+05 &  73.8 & 1926.978  6.85465e-05 &   0.7 &  -4.7 \\
      2002-09-26 07:01 & -11.26 & 4.04338E+05 &  50.5 & 1772.604  5.80036e-05 & 354.8 &  +1.3 \\
      2018-11-26 17:59 & -11.58 & 3.68917E+05 &  47.5 & 1942.799  6.96767e-05 &   0.5 &  +0.0 \\
      2018-04-02 09:27 & -12.07 & 3.88859E+05 &  23.8 & 1843.164  6.27133e-05 &   5.2 &  -5.6 \\
      2011-03-14 10:32 & -10.74 & 3.81174E+05 & -73.1 & 1880.327  6.52677e-05 & 351.2 &  +0.7 \\
      2017-12-02 12:49 & -12.47 & 3.63073E+05 & -15.6 & 1974.071  7.19378e-05 & 357.5 &  +6.5 \\
      2012-03-04 05:07 & -11.30 & 3.88740E+05 & -53.2 & 1843.731  6.27529e-05 & 353.9 &  +3.7 \\
      2011-11-16 23:42 & -10.79 & 3.87761E+05 &  70.6 & 1848.384  6.30690e-05 & 354.1 &  +4.8 \\
      2019-07-21 07:57 & -11.27 & 4.07689E+05 &  48.7 & 1758.035  5.70541e-05 & 357.9 &  +6.8 \\
      2015-02-07 05:12 & -11.71 & 4.08035E+05 &  34.7 & 1756.543  5.69573e-05 & 357.3 &  +2.9 \\
      \noalign{\smallskip}
      \hline
    \end{tabular}
    \end{center}
  \end{table*}
 
The phase angles of the Moon and its apparent diameter (in Table~\ref{tbl:moon_geom})
were calculated with the JPL HORIZONS
system\footnote{\url{https://ssd.jpl.nasa.gov/?horizons}}. It requires knowledge of the
Nadir Position, which is included in the level 1b data. The altitude of the orbit changes
just slightly during the mission and is available for each satellite from OSCAR.
Date and time are given for the mid-point of the Moon scan.
The central wavelength $\lambda_{chxx}$ for each infrared channel (and instrument),
with "xx" ranging from 01 to 19, are given at the project
web page\footnote{\url{https://www.nwpsaf.eu/site/software/rttov/download/coefficients/spectral-response-functions/}}.
Following the above-described
procedure, we extracted the counts from the warm and cold targets closest to the
Moon intrusion. Count errors are a combination of standard deviation of the counts
and digitization errors. Then the warm load blackbody (BB) temperatures (five temperatures for
HIRS/2 and HIRS/3 and six temperatures for HIRS/4) were converted to BB radiances
(Planck function applied to BB temperatures for wavenumbers\footnote{\url{https://www.nwpsaf.eu/site/software/rttov/download/coefficients/spectral-response-functions/}}).
We used the shifted relative spectral response functions and applied the
band-correction coefficients (correction coefficients from version 31 of the
file "calcoef.dat" in AAPP) to the central wavenumbers.
This radiance divided by the difference BB - DSV gave a slope.
The radiance of the Moon (in MJy/sr) was then calculated with the two-point calibration
using the slope and the difference Moon - DSV. This value is divided by the
fraction of the FOV filled by the Moon and the beam efficiency (see \citealt{Koenig1980}
and the User's Guide\footnote{\url{http://rain.atmos.colostate.edu/XCAL/docs/amsub/NOAA KLM Users Guide.pdf}}).
Its $\sigma$ was calculated with the error propagation
from the counts, assuming that the error of the BB Temp is zero.

\begin{figure}[h!tb]
\resizebox{\hsize}{!}{\includegraphics{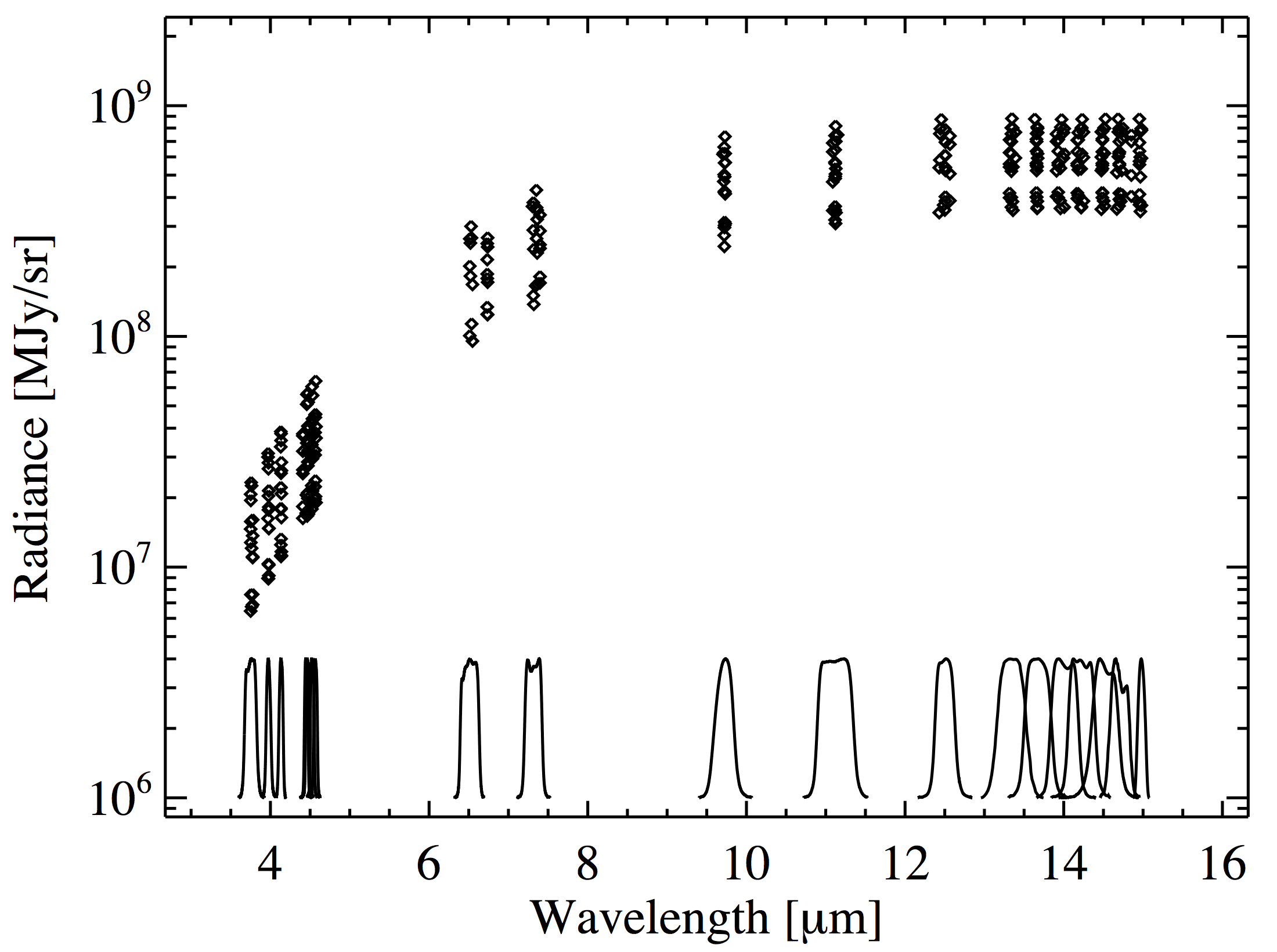}}
\caption{Calibrated Moon radiance values for the 19 HIRS thermal IR filters for all 22 observing
         epochs, as a function of wavelengths.
         The arbitrarily-scaled filter curves (NOAA-18/HIRS/4) are overplotted. The
         peak transmission of these narrow filters is always close to 1. The variation in
         radiance at a given wavelength is mainly related to the different heliocentric 
         distances of the Moon (ranging from r=0.985\,au to 1.018\,au) and the large
         range of phase angles (from -73.1$^{\circ}$ to +73.8$^{\circ}$).
         \label{fig:RadLamFil}}
\end{figure}

\begin{figure}[h!tb]
\resizebox{\hsize}{!}{\includegraphics{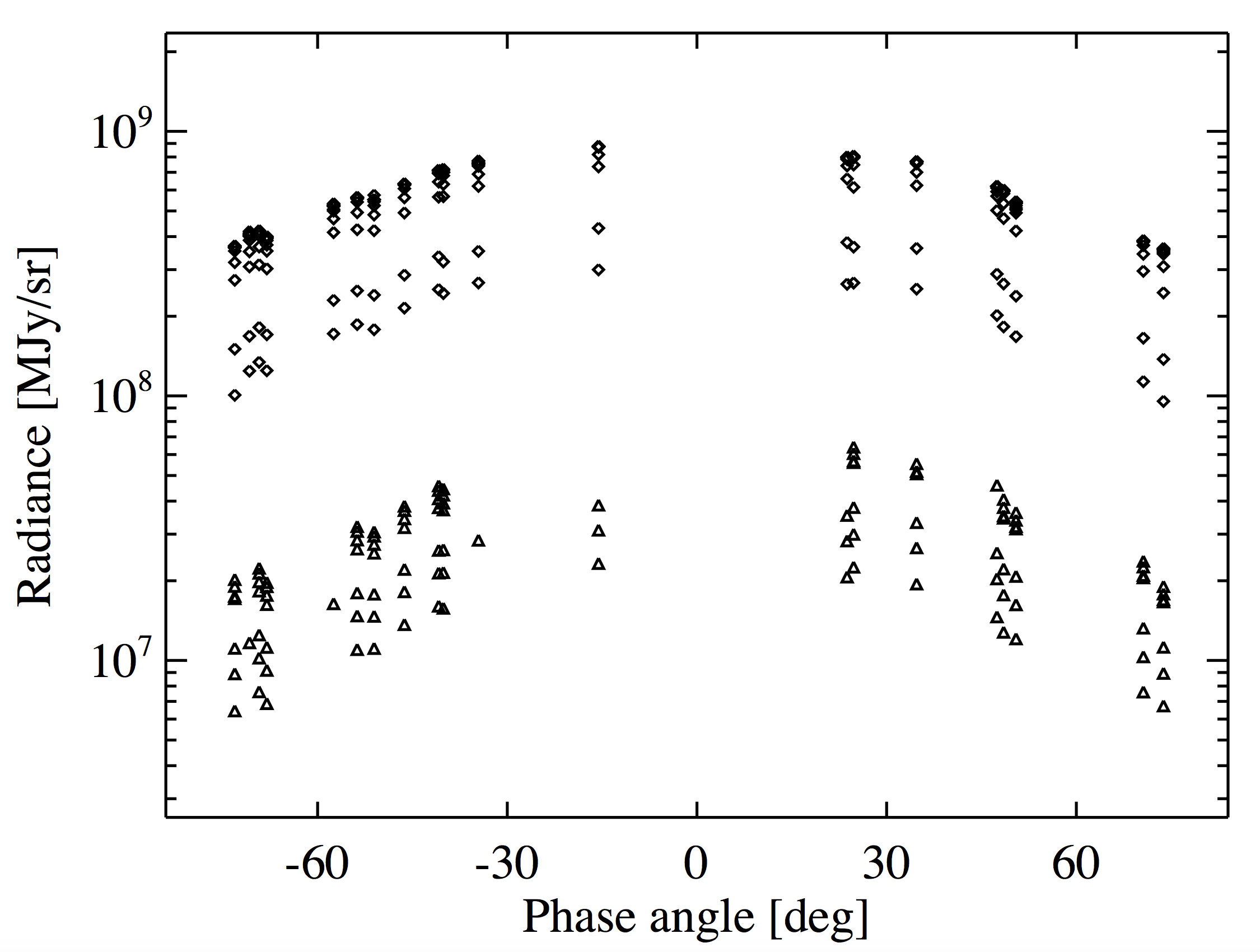}}
\caption{Calibrated Moon radiances for the 19 HIRS thermal IR filters for all 22 observing
         epochs, as a function of phase angle. The data at the top (diamonds) are from
         the long-wavelength channels 1-12 (14.9 to 6.7\,$\mu$m). The lower data points
         (triangles) are from the short-wavelength channels 13-19 (4.6 to 3.7\,$\mu$m).
         The variation in radiance at a given phase angle is mainly related to the
         different wavelengths (from below 4\,$\mu$m to almost 15\,$\mu$m) and heliocentric 
         distances of the Moon (ranging from r=0.985\,au to 1.018\,au).
         \label{fig:RadPhase}}
\end{figure}

For the direct comparison with TPM techniques, we convert the calibrated Moon radiances
(in MJy/sr) into (disk-integrated) flux densities (in Jy) by multiplying the radiances
with the apparent solid angle of the Moon at the times of the observations (column "diameter" 
in Table~\ref{tbl:moon_geom}).
We use the following conventions for the Moon's phase angles:
The phase angles for the waning Moon (indicated by "L" for "leading the Sun" in the
JPL HORIZONS system) have positive values, the phase angles for the waxing
Moon (indicated by "T" for "trailing the Sun" in the JPL HORIZONS system) have 
negative values, that is, the first quarter waxing Moon has $\alpha$=-90$^{\circ}$
and the third quarter waning Moon has $\alpha$=+90$^{\circ}$. Here, the
leading or trailing of the Sun is related to an observer placed on the given
satellite longitude, latitude, and altitude (see Table~\ref{tbl:hirs_events}) and in
consideration of the Earth's rotational direction.

The results from all 22 intrusions of
the Moon in the DSV that we identified in the raw data from various satellites are
compiled in Tables~\ref{tbl:hirs_obs_app_tbl1}, \ref{tbl:hirs_obs_app_tbl2}, and \ref{tbl:hirs_obs_app_tbl3}
(calibrated radiances and brightness temperatures), and in Tables~\ref{tbl:hirs_obs_app_long}
and \ref{tbl:hirs_obs_app_short} (flux densities). The random measurement errors in radiance values, brightness
temperatures, and flux densities are usually well below 1\%, and in very few cases close to 2\%. These 
errors are based on the scatter of multiple samples used for each radiance determination.
The calibrated Moon radiances are shown in Figs.~\ref{fig:RadLamFil} and \ref{fig:RadPhase}.

\section{TPM of the Moon}
\label{sec:tpm}

\subsection{TPM}

The interpretation of the measured fluxes and the comparison with model fluxes is done
via a thermophysical model (TPM) code developed by \citet{Lagerros1996I,Lagerros1997,Lagerros1998},
\citet{Mueller1998,Mueller2002a}, and \citet{Mueller2002}.
This model was used over the last two decades for near-Earth asteroids (NEAs)
\citep[e.g.,][]{Mueller2014b}, main-belt asteroids (MBAs) \citep[e.g.,][]{Ali-Lagoa2020}, 
satellites \citep[e.g.,][]{Detre2020}, or trans-Neptunian objects (TNOs)
\citep[e.g.,][]{Mueller2020}. It was extensively tested and validated against objects and
object properties that are known from direct measurements (i.e., occultations, radar,
or direct imaging) or from interplanetary mission rendezvous and flybys
\citep[e.g.,][]{Mueller2014,Mueller2014a}. Between 2016 and 2019, extensive efforts
were undertaken to compare quantities derived from TPM techniques with other methods and
to constrain the accuracy of TPM properties \citep{Mueller2018a}.
Overall, the TPM produced very high-quality diameters and albedos and consistent
thermal properties in cases where the available thermal data have good quality, cover
wide spans of time, wavelengths, and phase angles, which is why the new lunar HIRS
measurements are so valuable for our purposes.

\begin{figure}[h!tb]
\resizebox{\hsize}{!}{\includegraphics{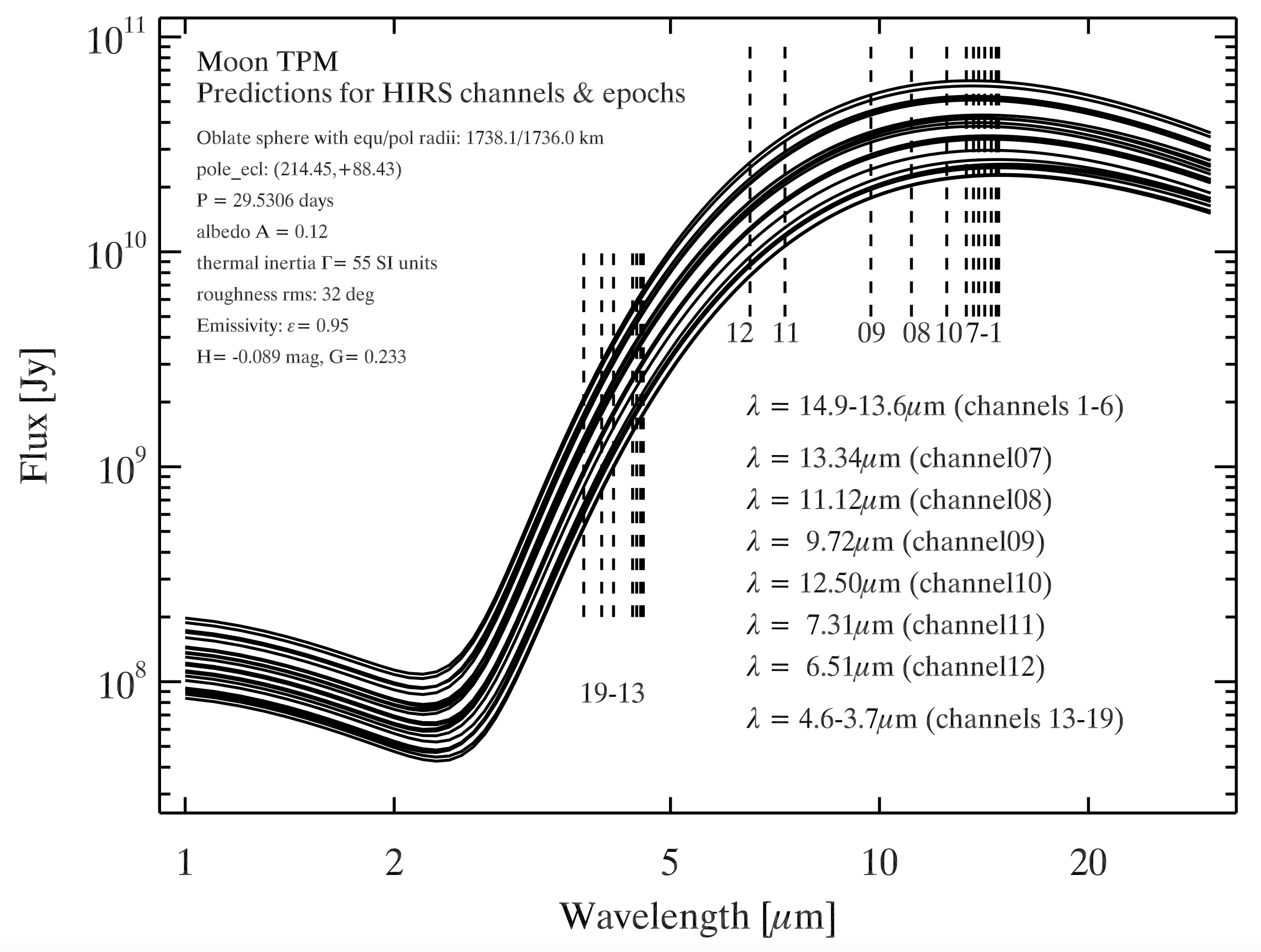}}
\caption{TPM predictions of the disk-integrated flux densities as a function of
         wavelengths for all 22 observing epochs. The central positions of the
         HIRS IR bands are indicated by dotted lines. The variation in absolute
         flux density is mainly
         related to the changing phase angles and, in second order, to the 
         small changes in heliocentric distance and observer-Moon distance. The
         thermal emission peak lies between 10 and 20\,$\mu$m, the transition
         between reflected sunlight and thermal emission is located between
         2 and 3\,$\mu$m. A constant hemispherical spectral emissivity of 0.95
         was assumed for these predictions.
         \label{fig:TpmSedPred}}
\end{figure}

The TPM predicts the thermal emission of atmosphereless bodies in the infrared to microwave
regime.
In the TPM the reflected sunlight can be estimated by using Lambert's scattering law,
but the calculations have not been optimized nor tested in the transition
region between reflected light and thermal emission where non-linear effects are involved.
The location and the width of the transition region depends on the object's heliocentric
distance, albedo, and thermal properties. For the Moon and near-Earth asteroids (NEA),
the thermal emission starts to
dominate beyond 3\,$\mu$m (see Fig.~\ref{fig:TpmSedPred}) for main-belt asteroids the
transition is located between 4-6\,$\mu$m; whereas
for more distant bodies, such as trans-Neptunian objects, the
transition happens at around 10\,$\mu$m or longer wavelengths.
The TPM can handle complex object shapes and it takes
the spin state into account. The heat conduction into the surface is calculated for
each surface facet. The surface roughness is modeled by hemispherical segments and
controlled by specifying the RMS of surface slopes \citep[see also][]{Davidsson2015}.
When calculating the directional- and wavelength-dependent emissivity, the TPM
considers sub-surface scattering processes.

The TPM can be used for radiometric studies to derive an object's size and albedo, as well as, in
some cases, thermal, shape, or rotational properties via the interpretation of
disk-integrated thermal measurements.
For objects with known properties, such as 
for the Moon or asteroids with in-situ studies, the TPM can make predictions of 
the object's surface temperatures, disk-integrated flux densities in the thermal
wavelength regime, or thermal lightcurves for aspherical bodies or objects with
albedo variations. These predictions can be used for a direct comparison with
thermal measurements, as here for the Moon or for calibration purposes
\citep[see e.g.,][]{Mueller2014}. 

The prediction of the thermal emission of the Moon can be described in three steps:
(i) the temperature of each surface element over the entire Moon has to be estimated;
(ii) the intensity $I_{\lambda}$ of each surface element in the direction towards the observer
has to be calculated; (iii) the disk-integrated flux is then determined as 
$F_{\lambda} = \frac{1}{\Delta^2} \oint I_{\lambda}(S) \mu dS,$
where $\Delta$ is the observer's distance, $dS$ is the surface element, and $I_{\lambda}$(S) is the corresponding
intensity towards the observer, and $\mu$ is the direction cosine, projecting the surface
element towards the observer. In the TPM, the $I_{\lambda}$ can be described as
$I_{\lambda} = \epsilon_d \frac{B_{\lambda}(\gamma T)}{\pi} + I_{sc} + I_{ref}$,
where T is the surface temperature and $B_{\lambda}$ is the Planck function.
$\gamma$ is the beaming function, $\epsilon_d$ is the direction and wavelength-dependent
emissivity ($\epsilon_d$/$\pi$ is the hemispherical spectral emissivity),
$I_{sc}$ is the multiple scattered radiation, and $I_{ref}$ is the reflected solar radiation.
For the temperature calculation, the Solar constant \citep[1361\,W\,m$^{-2}$ at 1\,au from the Sun;][]{Kopp2011},
the object's heliocentric distance, $r,$ and the surface albedo, A, are needed.
We estimated that the Earth's contribution to the energy balance of the Moon is negligible:
the thermal emission of the Earth is about 240\,W/m$^2$ (in all directions, averaged over
several years)\footnote{\url{https://science-edu.larc.nasa.gov/energy_budget/}}.
At the Moon distance, the Earth's contribution has to be scaled down by (6371\,km)$^2$/(384400\,km)$^2$,
which is less than 0.1\,W/m$^2$ \citep[see also][]{Glenar2019}.

The thermal emission of an idealized perfectly flat surface deviates from that of
a macroscopically rough one due to partial shadowing, the scattering of sunlight,
and self-heating. These effects lead to small areas with higher temperatures and more
thermal radiation being emitted in the direction back towards the Sun, known as the thermal-IR beaming effect.
There are different ways of describing and implementing surface roughness concepts
in TPMs \citep{Davidsson2015}. The most common way of expressing the thermally-relevant
level of surface roughness is by specifying the root-mean-square (rms) slope
\citep[see definition in][]{Spencer1990}.
We use a roughness implementation via hemispherical segmented craters where
the depth-to-diameter ratio and the surface crater coverage can be modified
to simulate different rms values and to calculate the beaming function,
$\gamma,$ \citep{Lagerros1996I,Lagerros1998}.
The $\gamma$ values are then used to correct the thermal emission for roughness effects.
At the end, the direction and wavelength-dependent emissivity, $\epsilon_d$, is required
for the prediction of the final (disk-integrated) flux, $F_{\lambda}$, at a given wavelength
\citep[see][]{Lagerros1996I,Lagerros1997,Lagerros1998,Mueller1998,Mueller2002a,Mueller2002}.

The energy balance on the surface is closely related to its albedo
and emissivity. It determines the daytime surface temperature of the Moon.
Since the Planck function is so strongly non-linear, it is important
to mention that the hottest surface regions dominate the disk-integrated emission
of an atmosphereless body at short wavelength up to the thermal emission peak.
Only at longer wavelengths, beyond the emission peak, the global, disk-averaged
object properties are thermally relevant.
The nighttime temperatures are tightly connected to the thermal inertia,
which is a function of thermal conductivity, density, and heat capacity. 
However, our HIRS measurements, taken between the first and the third quarter Moon,
are all dominated by the emission from the illuminated
part of the Moon and the thermal inertia is not important in this case.

\subsection{Relevant Moon properties}

For the Moon, we consider its equatorial radius of 1738.1 km, its
polar radius of 1736.0 km, its synodic period of 29.530589 days, and its
north pole coordinates with R.A.\ 266.86$^{\circ}$ and Dec.\ 65.64$^{\circ}$
(taken from the "Moon fact sheet"\footnote{\url{https://nssdc.gsfc.nasa.gov/planetary/factsheet/moonfact.html}},
\citet{Lang2012}, and from the IAU\footnote{Celest. Mech. Dyn. Astr.; DOI 10.1007/s10569-010-9320-4}).
The observing epoch and the satellite position, together with the Moon's pole axis, allow us
to calculate the sub-solar and sub-spacecraft longitudes and latitudes for each measurement,
namely, the true illumination and observing geometry. The sub-solar latitudes for our 22 epochs cover
the range from -1.6$^{\circ}$ to +1.5$^{\circ}$, while the sub-observer (spacecraft) latitude 
spans the range from -5.6$^{\circ}$ to +6.8$^{\circ}$.
\citet{Hayne2017} presented the global thermophysical properties of the Moon's regolith fines layer.
The diurnally active near-surface layer is about 4-7\,cm thick, with a top-layer thermal conductivity
of 7.4$\times$10$^{-4}$\,W\,m$^{-1}$\,K$^{-1}$ (3.4$\times$10$^{-3}$\,W\,m$^{-1}$\,K$^{-1}$
at a depth of $\approx$1\,m) and it has a globally averaged thermal inertia
$\Gamma$ = 55 $\pm$ 2\,J\,m$^{-2}$\,K$^{-1}$\,s$^{-1/2}$ at 273 K (and for
an albedo of 0.12).
Overall, the regolith fines are very uniform and lack hemispheric or any maria-highlands dichotomy.
Only small regions show significant deviations in
$\Gamma$: the interiors of prominent impact craters and specific crater ejecta have
$\Gamma$-values close to or above 100\,J\,m$^{-2}$\,K$^{-1}$\,s$^{-1/2}$ and the lunar cold spots
\citet{Bandfield2014} have $\Gamma$ $\le$50\,J\,m$^{-2}$\,K$^{-1}$\,s$^{-1/2}$.
Measurements at the equator show that the derived thermal inertia of a given area can vary
from around 35\,J\,m$^{-2}$\,K$^{-1}$\,s$^{-1/2}$ at midnight to about
70\,J\,m$^{-2}$\,K$^{-1}$\,s$^{-1/2}$ at noon due to the temperature dependence of the
thermal conductivity and heat capacity. However, for global flux calculations
in the HIRS wavelength and phase angle regimes, this temperature dependence of the thermal
inertia is not relevant since the observed flux is dominated by the hotter terrains that
are still visible.

The lunar average Bond albedo (at normal solar incidence) A is 0.12 \citep{Vasavada2012}. 
This is in agreement with the mean value of 0.122
found by \citet{Saari1972a}. \citet{Vasavada2012} derived a mean albedo of 0.07 for mare and 0.16 for
highland surfaces from measurements taken by the Diviner Lunar Radio Experiment.
In a NASA summary of the Moon's bulk parameters\footnote{\url{https://nssdc.gsfc.nasa.gov/planetary/factsheet/moonfact.html}},
the Bond albedo is given by 0.11 and the geometric albedo by 0.12.
The Moon's three-parameter magnitude phase function H, G$_1$, G$_2$
was determined by \citet{Muinonen2010}. The absolute (V-band) magnitude H= -0.154$^{-0.057}_{+0.118}$\,mag,
G$_1$ = 0.36$^{-0.12}_{+0.14}$, G$_2$= 0.338$^{-0.052}_{+0.049}$ fit the observed
reduced magnitudes over a wide phase angle range from 0 to 100$^{\circ}$ with
a residual rms of only 0.016\,mag, including the opposition surge of 0.43\,mag. The
two-parameter H-G$_{12}$ phase function values are H= -0.124$^{-0.020}_{+0.022}$\,mag,
G$_{12}$ = 0.358 $\pm$ 0.073. A simpler Lumme-Bowell fit to the phase function
by \citet{Bowell1989} gives H=-0.089\,mag and G=0.233. The phase integral q is 0.43 $\pm$ 0.04
\citep{Muinonen2010} or can be determined via $q = 0.290 + 0.684\,G = 0.45$ \citep{Bowell1989}.
The H-G, and q values play an important role in the context of radiometric asteroid
studies where size and albedo are derived from thermal measurements \citep[see e.g.,][]{Delbo2015}.
An asteroid's geometric V-band albedo p$_V$ can be calculated from its diameter D (often
derived from occultation measurements or given by radiometric solutions) and its absolute
magnitude H via $p_V = 10^{(6.2472-2\,ALOG10(D) - 0.4\,H_V)}$. To obtain the correct 
albedo value for the Moon, we have to use the Moon's size (3474.8\,km), together with
the H$_V$ magnitude but without the 0.43\,mag opposition surge.

The bolometric emissivity $\epsilon$ changes
with emission angle and surface roughness effects are also angle sensitive.
However, to be able to directly compare this study with the TPM framework used
for asteroids, we kept both properties fixed to their average values in our
global Moon model.
The bolometric infrared emissivity $\epsilon$ (as used for the temperature calculation)
was found to be 0.95, based on spectroscopic studies of lunar materials \citep{Donaldson2012}
and averaging over wide ranges of emission angles \citep{Bandfield2015}. This is the
apparent broadband hemispherical emissivity, originating from average daytime emission phase
function measurements.

The hemispherical lunar surface spectral emissivity ($\epsilon_d$ in the TPM convention)
can be assumed to be constant (with typical values of 0.90, 0.95, 1.00) or it can be
calculated from the measured reflectance values listed in the ECOSTRESS database\footnote{\url{https://speclib.jpl.nasa.gov/library}}:
There are 17 lunar reflectance spectra available, all derived from Apollo Moon samples
\citep[Apollo 11, 12, 14, and 16;][]{Salisbury1997}, and covering the wavelength range
from 2.079 to 14.011\,$\mu$m.
They include: (i) Apollo 11, 12 – Maria (lunar oceans): corresponding to "young soil" (darker material);
(ii) Apollo 16, (14) – Highlands: corresponding to "old soil" (brighter material);
(iii) Apollo 14 – Transitional case.
Overall, the emittance (1 - reflectance) spectral properties are dependent of the
surface type and show only subtle spectral variations between surfaces
of different composition and maturity \citep[e.g.,][]{Greenhagen2010}.
Only below 6\,$\mu$m do they differ substantially.

There are many studies and discussions in the literature that examine the relevant
roughness scales for the Moon and other atmosphereless bodies. \citet{Helfenstein1999}
produced digital topographic relief maps from closeup lunar images and measured
the surface roughness at 1-cm scale. They found a rms slope close to 10$^{\circ}$.
But since their image footprints were smaller than one square decimeter,
they had no information on the roughness statistics at meter or kilometer scales.
\citet{Ostro1993} derived the lunar rms slope to be $\approx$33$^{\circ}$. His radar
observations were sensitive to spatial scales between the radar wavelength and the
spot size of the sub-radar point. \citet{Rozitis2011} discussed a wide range of
lunar roughness studies and applied their model to lunar measurements presented
by \citet{Saari1972a}. In the summary of their work, they listed lunar surface roughness
values between 30$^{\circ}$ and 39$^{\circ}$, all derived by thermal models.
Their own best value is $\approx$32$^{\circ}$, inferred from the lunar thermal-IR
beaming effect, and in agreement with the radar studies by \citet{Ostro1993}.
\citet{Rosenburg2011} quantified the surface roughness properties of the Moon
based on data from the Lunar Orbiter Laser Altimeter (LOLA), carried on the
LRO. Using baselines ranging from $\approx$17\,m to several kilometers, they
mapped the lunar surface in different roughness parameters and they found
vastly different roughness properties for the lunar highlands and the mare plains.
\citet{Bandfield2015} found that Nadir observations can be modeled using
20$^{\circ}$ RMS slope distribution, while multiple emission angle observations
are best matched by a 20-35$^{\circ}$ distributions. Their data did not show any clear
variation in roughness between different regions or surface units.
However, \citet{Rubanenko2020} used telescope \citep{Sinton1962} and LRO Diviner data
\citep{Bandfield2015} to produce a detailed lunar surface roughness map. They
list 30.2$^{\circ}$ $\pm$ 5.9$^{\circ}$ (Maria) and 36.8$^{\circ}$ $\pm$ 4.4$^{\circ}$
(Highlands) as representative values for the bidirectional RMS slope at the thermal
insolation scale, consistent with results presented by \citet{Bandfield2015, Rozitis2011}.
In the context of our study of disk-integrated thermal measurements of the Moon,
it is important to note that the roughness properties derived
from baselines up to the multiple-kilometer range dominates the measured
thermal-IR beaming effects of the unresolved Moon observed by the HIRS instrument
\citep[see also][]{Rozitis2011}.

\section{TPM Moon predictions \& HIRS data}
\label{sec:comp}

Based on the given sun illumination and observing geometries (see Tables~\ref{tbl:hirs_events}
and \ref{tbl:moon_geom}) and the above-listed size, shape, and spin properties, we made TPM
flux density predictions for a direct comparison with the measurements.
The slow rotation combined with low-conductivity surface layers dramatically
reduce the influence of the thermal inertia on the day-time temperatures. 
Therefore, modifications of this value to account for temperature variations or
regional differences are not needed and, thus, we keep it fixed at 55\,J\,m$^{-2}$\,K$^{-1}$\,s$^{-1/2}$.   
The crucial properties in our study are albedo, emissivity, and surface roughness.

\subsection{Influence of roughness}

\begin{figure}[h!tb]
\resizebox{\hsize}{!}{\rotatebox{0}{\includegraphics{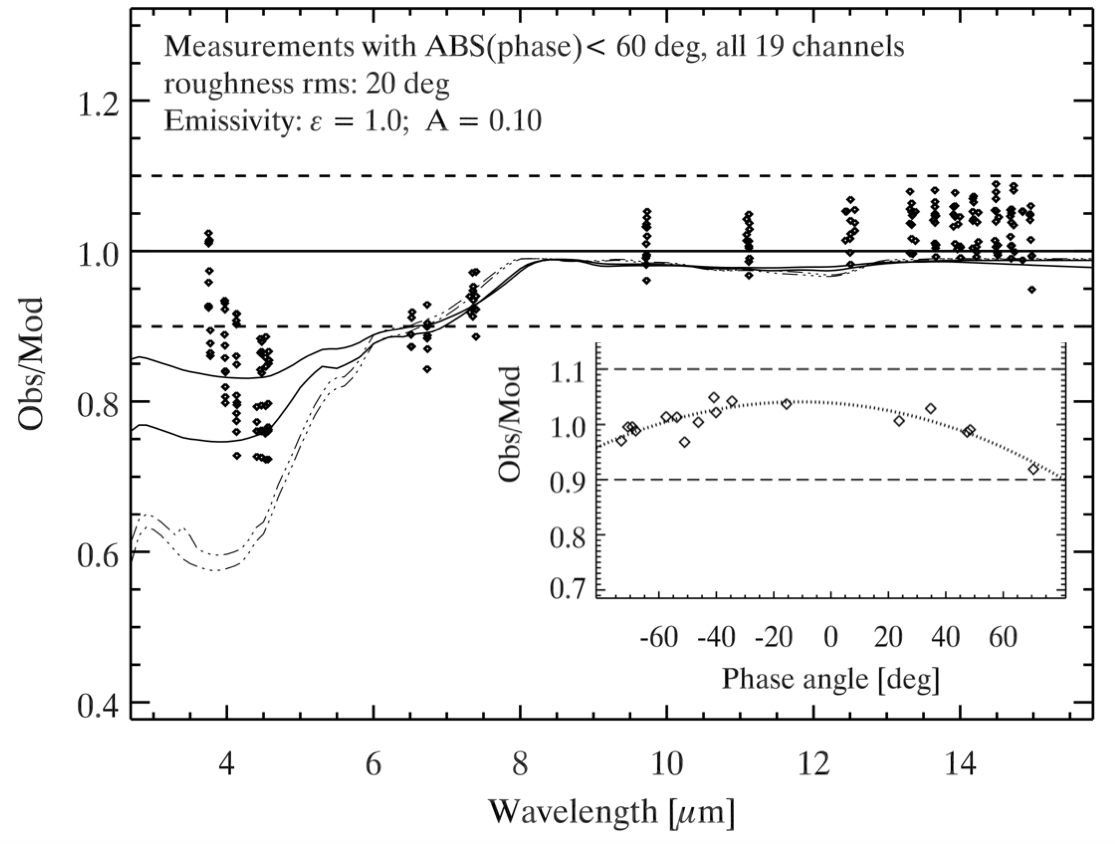}}}
\resizebox{\hsize}{!}{\rotatebox{0}{\includegraphics{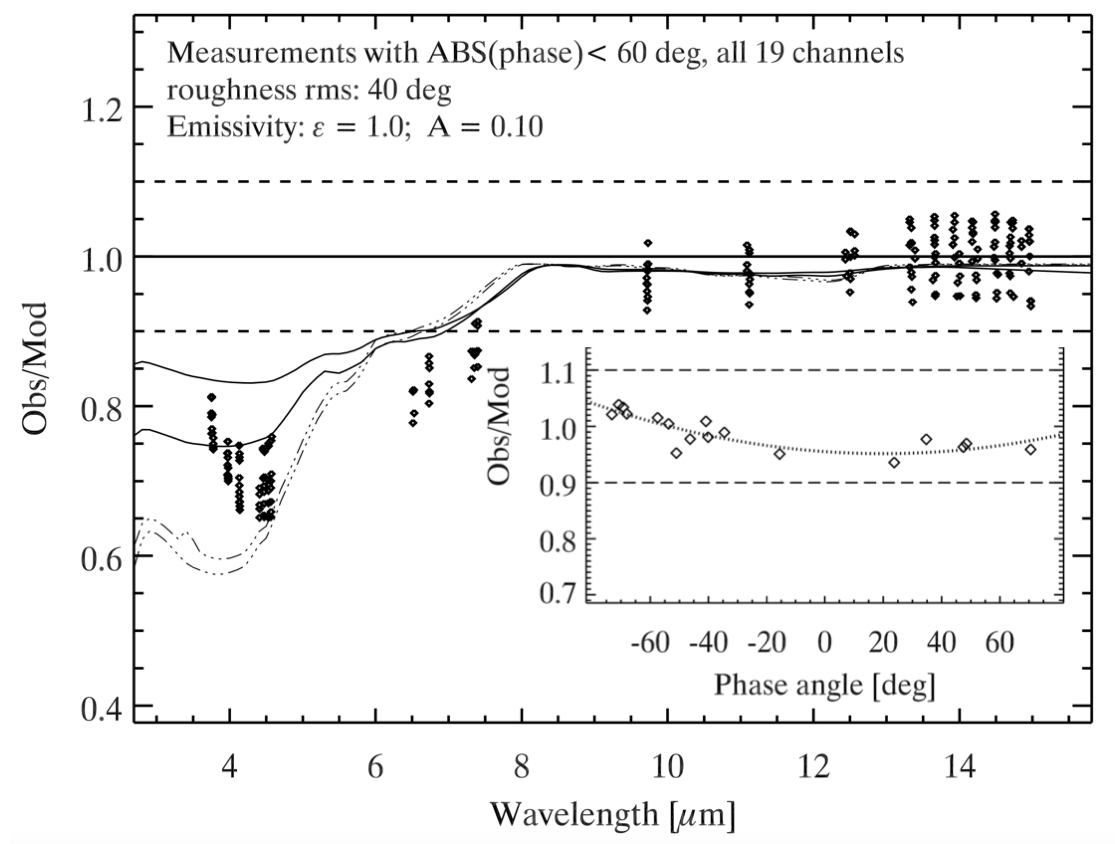}}}
\caption{Observations divided by TPM predictions ($\epsilon = 1.0$; albedo A=0.10) as a function of wavelengths
         (and for channel 08 at 11.1\,$\mu$m also as a function of phase angle in the inserted figures).
         The measurements are shown as diamond symbols, the scatter within each channel or wavelength
         is associated to a small but systematic deviation dependent on phase angle.
         The ECOSTRESS emissivity (calculated as 1-reflectance) spectra for two extreme maria samples
         (solid lines) and two extreme highland samples (dotted-dashed lines) are shown.
         All other (maria, transition \& highland) samples would fall between these lines.
         Top: For a low surface roughness (rms 20$^{\circ}$). Bottom: Using a 
         high roughness (rms 40$^{\circ}$) in the TPM. The dotted lines in the phase angle plots are second
         order fits through the ratios.
         \label{fig:RoughnessTest}}
\end{figure}

The greatest influence (or uncertainty) in our TPM flux prediction comes
from the surface roughness. In a first test, we set the model emissivity values
to 1.0 at all HIRS wavelengths. This setting enables a direct comparison
with the measured spectral emissivity properties for multiple lunar
samples \citep{Salisbury1997}. We also did not account for the
reflected sunlight (see spectral energy distribution (SED) part below $\approx$2.5\,$\mu$m in
Fig.~\ref{fig:TpmSedPred}).
Figures~\ref{fig:RadLamFil} and \ref{fig:RadPhase} show the measured and calibrated
absolute radiances versus\ wavelengths and phase angle, respectively.
Figure~\ref{fig:RoughnessTest} shows all HIRS
measurements, taken between -60$^{\circ}$ and +60$^{\circ}$ phase
angles\footnote{We excluded the NOAA-17 data from 2002 taken at $\alpha$=50.5$^{\circ}$
where the Moon might have been partly outside the FOV at the longest wavelengths.},
divided by TPM predictions, assuming a constant albedo of 0.1.
ECOSTRESS spectra (calculated as 1 - reflectance) of two lunar
mare samples (solid lines), while two highland
spectra (dotted-dashed lines) are overplotted to guide the eye.
At short wavelengths below 10\,$\mu$m, the lower roughness with a rms slope
of 20$^{\circ}$ (top part of Fig.~\ref{fig:RoughnessTest}) pushes
the ratios to a good match with dark maria emissivity spectra. At very short
wavelength (below 5\,$\mu$m) the influence of reflected sunlight becomes apparent
and the measured fluxes exceed the characteristic lunar spectra.
A higher surface roughness with a rms slope of 40$^{\circ}$ (bottom part
of Fig.~\ref{fig:RoughnessTest}) increases the TPM fluxes, hence, the displayed
ratios go down. The high roughness matches the long-wavelength ($\lambda$
$>$10\,$\mu$m) very well, but does not follow any of the lab spectra in the
6-10\,$\mu$m range. The roughness has also an effect on the phase curves.
A low roughness in the model setup underestimates the true fluxes close to full Moon and 
overestimates the ones at large phase angle. For high levels of surface roughness, 
we see the opposite. In Figure~\ref{fig:RoughnessTest} (inserted plots), we show this
effect for
our two extreme roughness levels (rms of 20$^{\circ}$ and 40$^{\circ}$) for channel 08
at 11.1\,$\mu$m. The dotted lines are second-order fits to the resulting ratios.
The other channels show a similar behavior. At an intermediate roughness level (best solution
is at $\approx$32$^{\circ}$), the observed phase curves are matched by the TPM predictions
and the fits through the ratios are flattened out.

None of the different surface roughness levels bring the TPM predictions
into agreement at all wavelengths with the available ECOSTRESS lunar
emissivities for the Apollo samples. If we assume that the
lunar maria spectra are more relevant (darker zones are hotter and contribute
more to the thermal emission at these wavelengths) then this would point to
a strongly wavelength-dependent surface roughness (low roughness values at
short wavelengths and high values at longer wavelengths) that is unphysical.
The emission measured by HIRS originates from the very top few millimeters
to centimeters of the surface (the thermal skin depth is $\approx$1\,cm)
and baselines up to several kilometers play a role for the 
relevant roughness properties \citep[see also][]{Rosenburg2011,Rozitis2011,Rubanenko2020}.
\citet{Rozitis2011} summarized lunar roughness studies on different scales,
including results from radar measurements.
They analyzed 10-12\,$\mu$m scans
of the sunlit portion of the Moon obtained by \citet{Saari1972a} and found
that rms slopes close to 32$^{\circ}$ explain the lunar IR beaming effect
in the most consistent way. Our best-fit roughness solution
confirms this value and we use
it as default from now on for all HIRS channels.

\subsection{Global versus\ local albedo}

In a second test, we set the model emissivity values again to 1.0.
For the surface roughness we use the best-fit rms slope of
32$^{\circ}$ and we only modify the albedo.
In principle, the global average of A=0.12 is known, but we tested
for lower and higher values. We used an albedo of 0.07 (average maria value)
and 0.16 (average highland value) in the model calculations.

The effects of different albedo values are slightly smaller than the
ones found for roughness and the comparison with the measurements
looks very similar to the ratios shown in Figure~\ref{fig:RoughnessTest}
(low albedo has a similar effect as high roughness and vice versa).
The thermal IR fluxes at short wavelengths are modeled best by
a low, mare-like albedo. At the long HIRS wavelengths the measurements
are in better agreement with an intermediate albedo value around 0.1.
Very high albedo values (0.16) underestimate the long-wavelength data
and the observation-to-model ratios systematically exceed 1.0.

The albedo signature in the data is clear: the darkest
terrains on the surface are the hottest and the highest temperatures dominate
the observed fluxes at short wavelength. At longer wavelengths, closer to the
thermal emission peak, more and more lower-temperature zones contribute to the
disk-integrated fluxes, closely related to a global average albedo value. Since
our TPM setup uses only a single albedo for the entire Moon, we took, similarly
to the work by \citet{Rozitis2011}, a value of A=0.10.

\subsection{Emissivity}

\begin{figure}[h!tb]
\resizebox{\hsize}{!}{\rotatebox{0}{\includegraphics{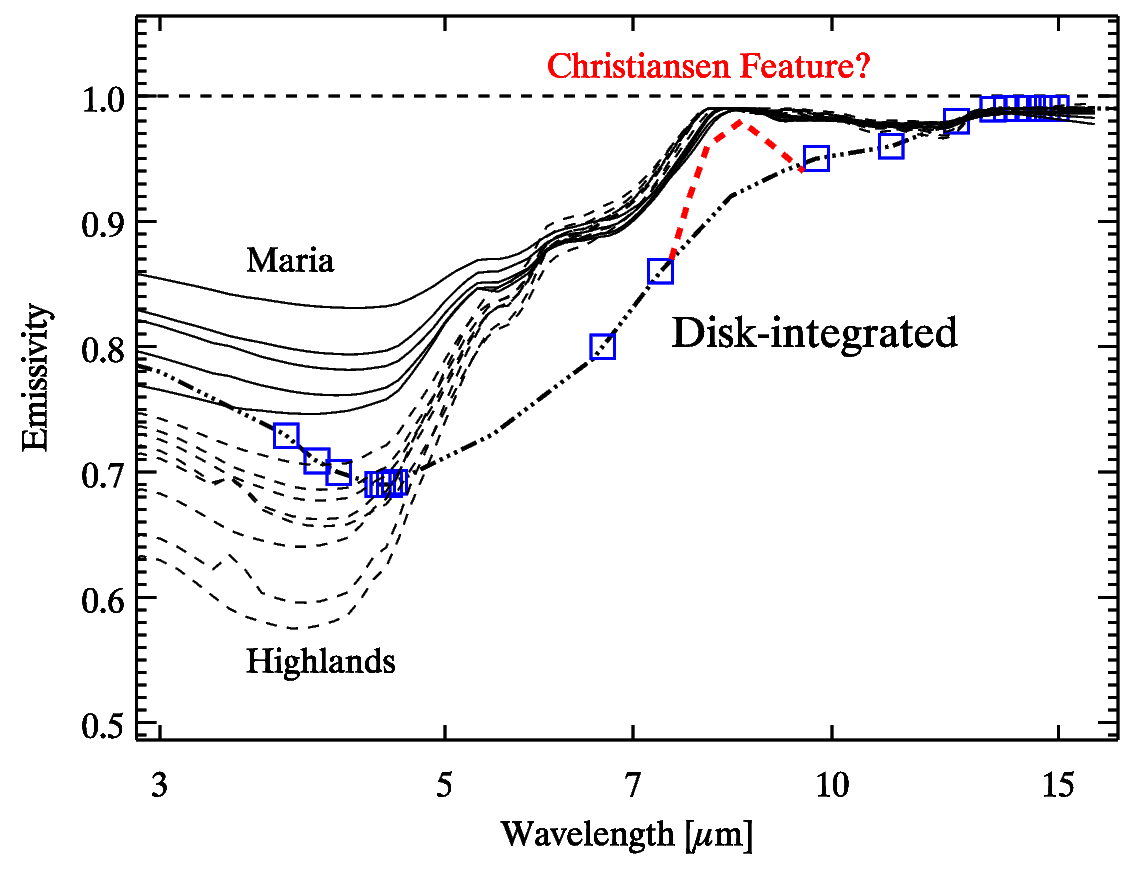}}}
\caption{Available maria (solid lines) and highland (dashed lines) ECOSTRESS
         lunar hemispherical emissivities together with our emissivity spectrum
         (dotted-dashed line) derived from global, disk-integrated
         Moon measurements (blue square boxes).
         Our model emissivity is not very well constrained at wavelength below 4\,$\mu$m
         (due to reflected sunlight contributions) and in the ranges 4.6-6.5\,$\mu$m and
         7.3-9.7\,$\mu$m (due to the lack of measurements).
         \label{fig:EmiModels}}
\end{figure}

One of the goals was the production of a lunar TPM setup that explains
the HIRS data with the best possible accuracy. Therefore, we had to establish
our own spectral emissivity model from the HIRS data. Once again, we produced the
observation-to-model ratios with our default roughness (rms =32$^{\circ}$)
and albedo (A=0.10) values and assuming a constant flat emissivity of 1.0.
This is shown in Fig.~\ref{fig:EmiModels},
together with the available lunar mare and highland spectra. There are five
additional "transitional" samples that are not shown, but they look very
similar and lie in the middle between these dark and bright materials.
Our new global lunar emissivity model was established by a fit through
all HIRS channels (dashed-dotted line in Fig.~\ref{fig:EmiModels}).

It is interesting to see that the global hemispherical emissivity as
derived from our HIRS measurements roughly agrees with the
averaged maria and highland spectra, both at short wavelengths below 5\,$\mu$m
and also at long wavelengths beyond 11\,$\mu$m. At intermediate wavelengths between
5 and 11\,$\mu$m (HIRS channels 12, 11, 09, at 6.5\,$\mu$m, 7.3\,$\mu$m, and 9.7\,$\mu$m,
respectively), the measured effective global emissivity
differs clearly from the lunar sample emissivities \citep{Salisbury1997}.
There are no indications from the HIRS calibration activities that these three channels
have any calibration issues.
Another important aspect is that at wavelengths between the HIRS
channels, we have no information on the hemispherical emissivity.
The regions between 4.6 and 6.5\,$\mu$m and between 7.5 and 9.5\,$\mu$m stand as
the most prominent gaps. Here, the true values are not constrained by our measurements.
In particular, one problematic region for our model is located in the range of the Christiansen
feature \citep[e.g.,][]{Murcray1970,Greenhagen2010}. This feature is
located between $\approx$7.5 and 9.0\,$\mu$m (indicated by the red
dashed line in Fig.~\ref{fig:EmiModels}). \citet{Vasavada2012} found peak
emissivities of 0.96 to 0.98 in the 8 to 8.5\,$\mu$m range, related to
specific locations on the lunar surface. This work was based on the
three mineralogy channels (A3: 7.80$\pm$0.25\,$\mu$m; A4: 8.20$\pm$0.22\,$\mu$m;
A5: 8.60$\pm$0.20\,$\mu$m) of the Diviner Lunar Radiometer Experiment.
The HIRS channels have no overlap with the Diviner channels and have very narrow
filters (see Fig.~\ref{fig:RadLamFil}). Therefore, they cannot be used to
constrain the global emissivity in the range of the Christiansen feature.
Another aspect which limits our comparison is that the lunar samples have been
measured under particular temperatures and illumination or observing angles,
while the HIRS-derived emissivities are the result of combined multi-angle and
multi-temperature conditions on the surface of the Moon. But the dominating reason
for the discrepancy is not clear. Further full-disk measurements are needed to
confirm our findings and to fill the gaps between the HIRS channels in order to achieve a
characterization of the global emissivity.

\begin{figure}[h!tb]
\resizebox{\hsize}{!}{\rotatebox{0}{\includegraphics{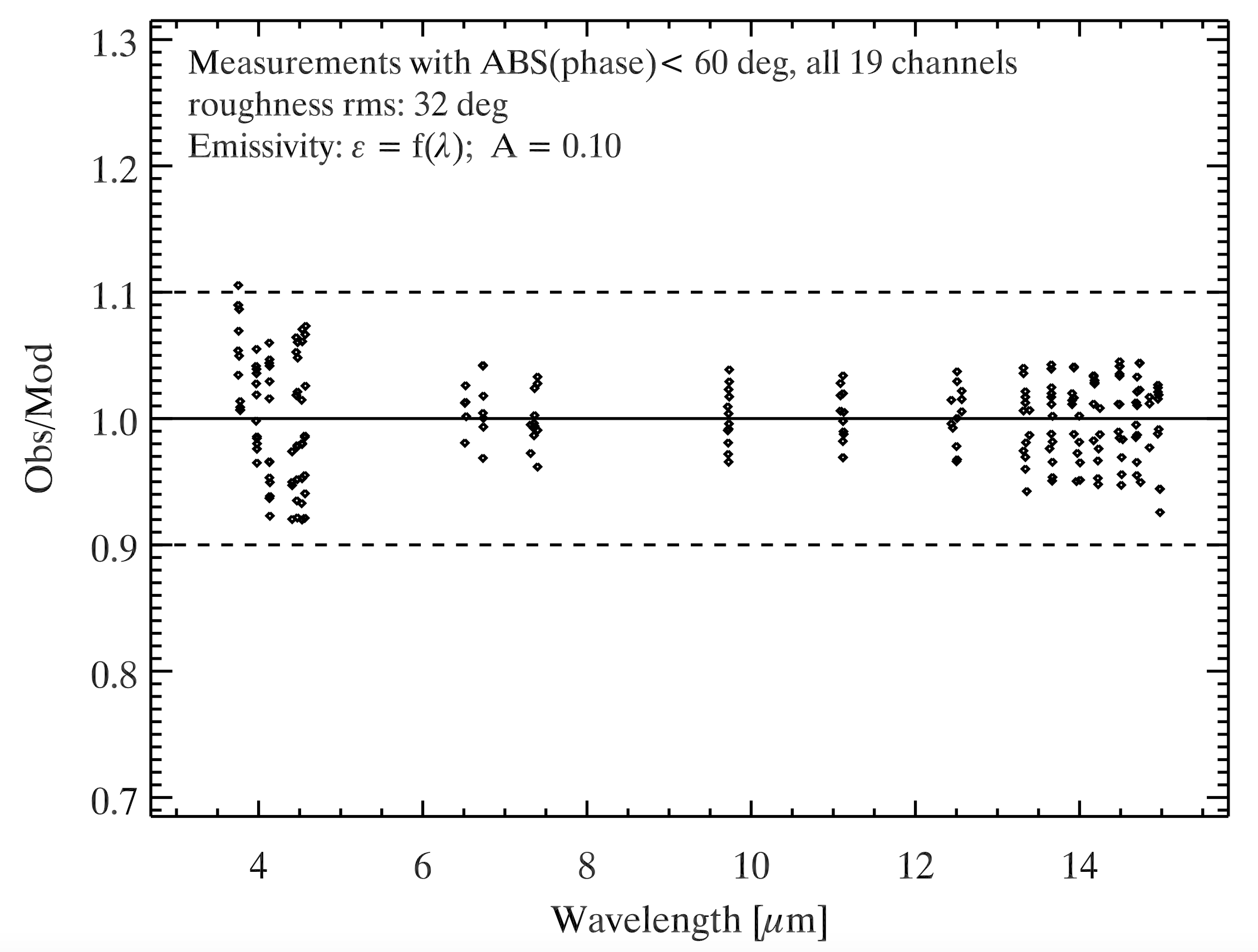}}}
\caption{Observations divided by TPM predictions, using the default rms slopes of
         32$^{\circ}$ and an albedo A of 0.1, but now with a hemispherical spectral
         emissivity derived from the corresponding observation-to-model ratios.
         The reflected light contribution is again visible as an increase of the ratios
         at the shortest wavelengths.
         \label{fig:EmissivityTest}}
\end{figure}

Using our new "lunar global emissivity spectrum", the ratios are brought close to
1.0 (see Figure~\ref{fig:EmissivityTest}) while the scatter is reduced at a given
wavelength or phase angle to a minimum. Beyond 5\,$\mu$m, the TPM predictions
agree now within 5\% of the measured values, while at shorter wavelength, we are still
within 10\%. Outliers are found in ch18/ch19 at very short wavelengths, where the
reflected sunlight contributes a few percent to the measured values, and at the
longest wavelengths, where the noise levels are higher and where it was not
always perfectly clear whether the Moon was completely in the FOV.

Temperature gradients can be extremely steep in the upper few millimeters of 
the lunar surface \cite[e.g.,][]{Keihm1984,Bandfield2015}, and, in the beginning,
it was not clear if the different spectral channel would be sensitive to different
sub-surface layers. The emission measured by HIRS originates
from the very top few millimeters to centimeters of the surface.
At these short wavelengths, the thermal emission is
dominated by the hottest temperatures on the surface
and the sub-surface would not contribute significantly
to the total disk-integrated flux.
This can also be seen in Fig.~\ref{fig:EmissivityTest}. 

By applying our
global spectral emissivity solution, which shows a similar behavior as the
Apollo sample emissivities measured at a constant temperature, we can fit the
HIRS measurements over all channels equally well. There are no indications that
the different spectral channels have a depth sensitivity. In addition,
asteroid thermal emission studies show that sub-surface layers seem to 
influence the disk-integrated fluxes only starting in the far-IR, sub-millimeter
or millimeter range \citep[see e.g.,][]{Mueller2014}.

\subsection{Phase curves} \label{sec:phasecurves}

Our HIRS data cover a phase angle range from -73.1$^{\circ}$ (waxing Moon)
to +73.8$^{\circ}$ (waning Moon). We consider whether the Moon phase curves at
thermal wavelengths are symmetric with respect to opposition (phase angle $\alpha$=0$^{\circ}$)
and whether the phase slopes are well explained by our TPM solutions.
\citet{Maghrabi2014} obtained groundbased 8-14\,$\mu$m measurements over a full
Moon cycle. They found
that the IR temperature reaches its maximum of 391 $\pm$ 2\,K for the full Moon
and claim a symmetric phase behavior. They also determined a temperature
of 240 $\pm$ 3.5\,K for the first quarter, and 236 $\pm$ 3\,K for the last quarter.

\begin{figure}[h!tb]
\resizebox{\hsize}{!}{\rotatebox{0}{\includegraphics{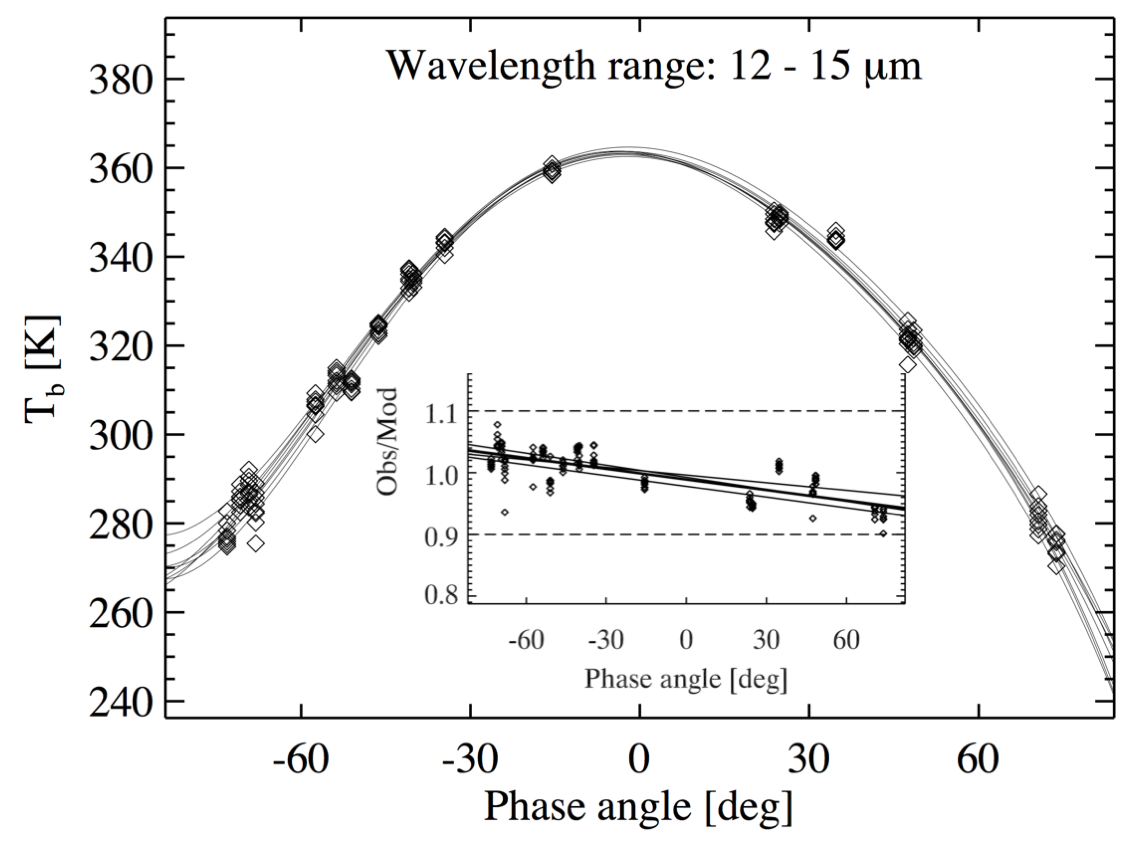}}}
\resizebox{\hsize}{!}{\rotatebox{0}{\includegraphics{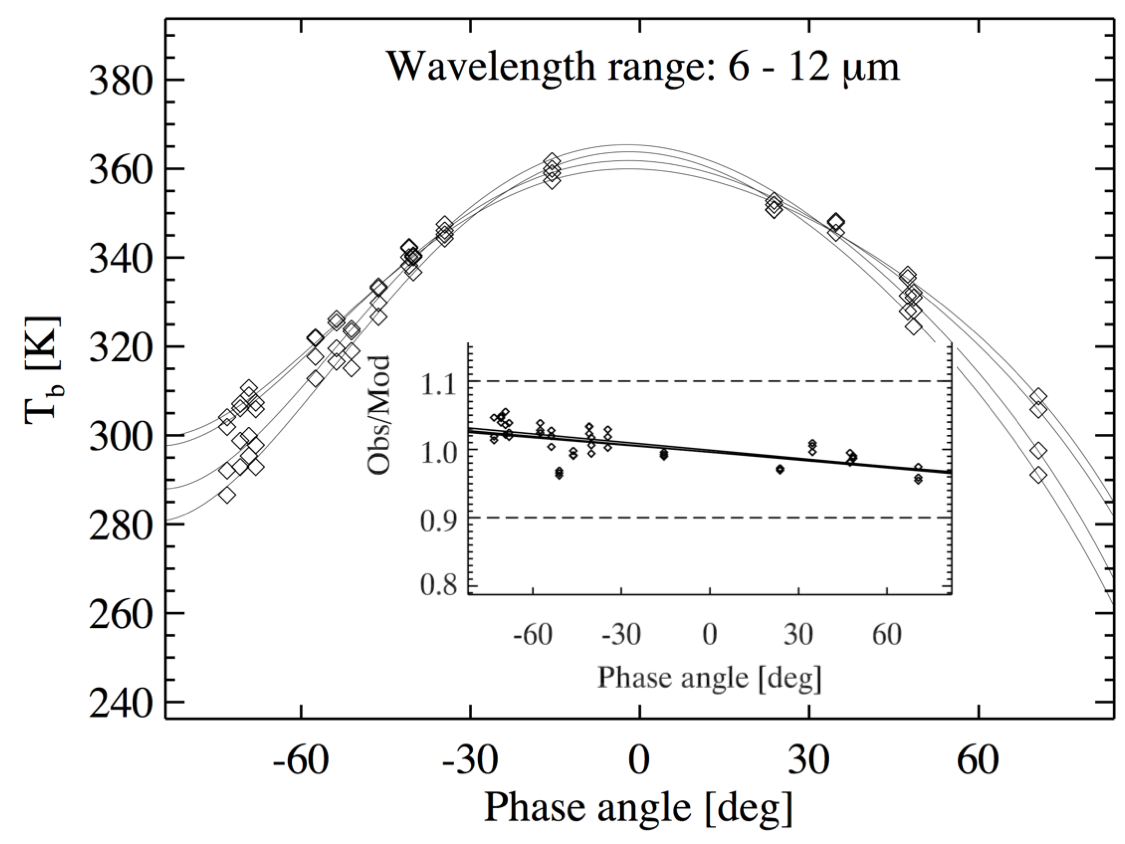}}}
\resizebox{\hsize}{!}{\rotatebox{0}{\includegraphics{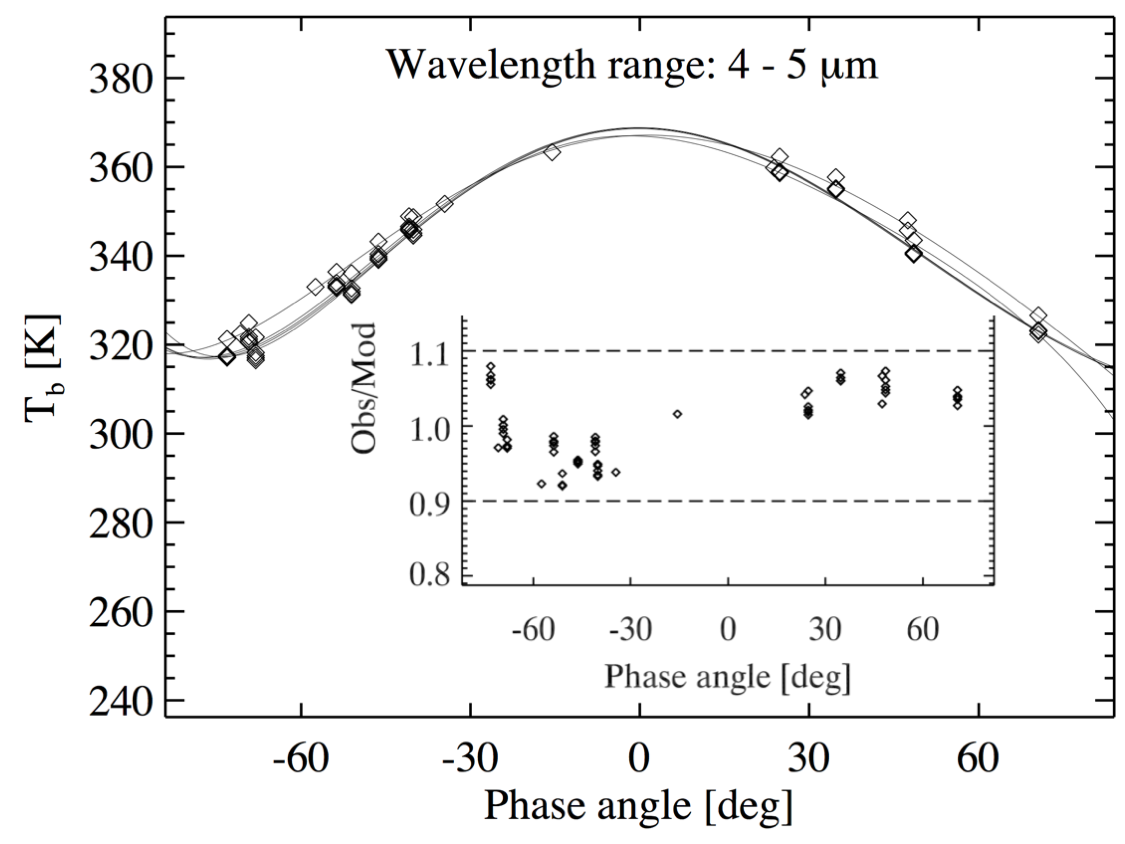}}}
\caption{Measured and calibrated HIRS brightness temperatures as a function of
         phase angle and fitted (in each channel separately) by a fifth-order
         polynomial. Top: For long-wavelength channels (12-15\,$\mu$m); Middle:
         For  mid-wavelength channels (6-12\,$\mu$m); Bottom: For short-wavelength
         channels (4-5\,$\mu$m). Phase curves are asymmetric and with a 2-3$^{\circ}$
         shifted peak (towards negative phase angles) at mid to long wavelengths.
         The inserted plots show the corresponding flux densities divided by TPM
         predictions, also as a function of phase angle. The lines have
         been fitted in each channel separately. The before-after opposition
         asymmetry is clearly visible for the long- and mid-wavelengths channels.
         At short wavelengths, there is a wavy structure and no linear fit was done.
         \label{fig:BrightnessTempPhase}}
\end{figure}

We also looked at the HIRS brightness temperatures as a function of phase angle, but
for each of the HIRS channels separately. Figure~\ref{fig:BrightnessTempPhase}
shows (from top to bottom) the data from HIRS long-wavelength channels (12-15\,$\mu$m),
mid-wavelength channels (6-12\,$\mu$m), and the short-wavelength channels (4-5\,$\mu$m).
Channels 18 and 19 are affected by reflected sunlight and have been excluded.
We fit each channel separately with a fifth-order polynomial, as done for the microwave
range (89 to 190\,GHz) by \citet{Burgdorf2019}. These phase curves show
slightly asymmetric shapes. In the short-wavelength 4-5\,$\mu$m range the fitted
peak brightness temperatures (368.0 $\pm$ 0.9\,K) are found close to opposition at
0.2$^{\circ}$ $\pm$1.2$^{\circ}$. In the LW channels, the peak brightness temperatures
(363.2 $\pm$ 1.4\,K) are consistently shifted by 2-3$^{\circ}$ towards negative
phase angles (-2.6$^{\circ}$ $\pm$ 0.7$^{\circ}$), that is, a few hours before full Moon.
In the LW data there is even a trend for the brightness peak to move to more negative
phase angles at longer wavelengths.

At much longer wavelengths in the microwave regime, \citet{Keihm1984}, \citet{Burgdorf2019}
or \citet{Liu2020}
found a significant phase shift of the diurnal maximum.
The peak brightness temperature maximum occurs at a phase angle
of about 20$^{\circ}$ and 24$^{\circ}$ at 157\,GHz and 89\,GHz, respectively.
The lag angle
depends on the ratio between the physical thickness of the emission layer and the penetration
depth \citep{Krotikov1964}. However, at the HIRS wavelengths, we only see the millimeter-to-centimeter
top-layer of the surface and the shift to negative phase angles is very likely related
to small hemispherical differences in thermal properties (roughness and albedo), which sum
up to a slightly shifted and asymmetric phase curve.

The phase curves are also relevant in the context of TPM concepts to see how well
the flux predictions match the observed disk-integrated flux densities of the Moon.
We take the best TPM solutions for the short- and long-wavelength HIRS regime,
calculate again the ratios between observed and modeled flux densities, and 
show the result as a function of phase angle (see inserted plots in Fig.~\ref{fig:BrightnessTempPhase}).
The model includes now our disk-integrated hemispherical emissivity model and is
calculated for an albedo of 0.10, roughness rms slopes of 32$^{\circ}$,
and the above-listed in-situ properties. The asymmetry in the observation-to-model
ratios can be corrected via the average fitted slope. In the 12-15\,$\mu$m range
the TPM predictions have to be increased by 0.56$\pm$0.07\% per 10$^{\circ}$ 
phase angle for the waxing Moon and decreased for the waning Moon. 
In the 5-12\,$\mu$m range the correction is smaller with only 0.35$\pm$0.06\% per 10$^{\circ}$
phase angle.
At short wavelengths below 5\,$\mu$m, the corrections are not as well defined (see
inserted plot in Fig.~\ref{fig:BrightnessTempPhase}, bottom). The wave-like 
sinusoidal pattern could give a first hint for possible corrections, but more
data would be needed to quantify these corrections in a meaningful way.

The before-after opposition asymmetries in the residuals between observations
and TPM predictions are very likely
caused by temperature differences between morning and afternoon. At 
positive phase angles, the HIRS measurements sample mainly "morning" illuminated
slopes, whereas the negative phase angles sample more afternoon-evening slopes,
which are slightly hotter than the morning ones. Additional contributions might also
be related to albedo \citep{Vasavada2012} or roughness properties
\citep{Rubanenko2020} that show regional variations over the lunar surface.
At large negative phase angles the illuminated fraction of the surface has 
probably a lower mean albedo, leading to higher temperatures and a few percent
higher (than predicted) fluxes and vice versa. At the shortest wavelengths
(inserted plot in Fig.~\ref{fig:BrightnessTempPhase}, bottom), the trend is probably
caused by small dark (or not so dark) areas that dominate the
full-disk lunar emission at specific phase angles. Global mean values of albedo
and roughness are less relevant.

After fitting the phase asymmetry (inserted plots in Fig.~\ref{fig:BrightnessTempPhase}) in
the LW channels, we find remaining deviations from the fitted lines
of only 1-3\% in a given channel. The best agreement between the HIRS
data and our final TPM predictions are found for channels 8-12 (6.5-12.5\,$\mu$m).
In the SW channels, where fitting of the phase asymmetry was not done, the
individual measurements can deviate up to 8\% from our TPM predictions.

\clearpage
\section{Applications and discussion}
\label{sec:dis}

We find an excellent agreement between the HIRS measurements of the Moon
and our TPM predictions when using the following model parameters:
(i) the physical size, shape, rotational properties of the Moon;
(ii) a thermal inertia $\Gamma$ = 55\,J\,m$^{-2}$\,K$^{-1}$\,s$^{-1/2}$
     \citep{Hayne2017};
(iii) a surface roughness characterized by an rms slope of 32$^{\circ}$
      (as in \citet{Rozitis2011});
(iv) an albedo A= 0.10;
(v) reflected light properties following the Lumme-Bowell convention:
    H=-0.089\,mag (with 0.43\,mag opposition surge), G=0.233, q= 0.43
    (not crucial for thermal emission calculations);
(vi) our lunar global hemispherical spectral emissivity model
     (see Fig.~\ref{fig:EmiModels}); and, (vii) phase-angle asymmetry model
     corrections of up to +3\% at $\alpha$ $<$-60$^{\circ}$
     and -3\% at $\alpha$ $>$+60$^{\circ}$, but only for wavelengths $>$6\,$\mu$m
     (see inserted plots in Fig.~\ref{fig:BrightnessTempPhase}, top \& middle;
     corrections given in the text).

This model is valid in the wavelength ranges 5-7.5\,$\mu$m and 9.5-15\,$\mu$m
and for phase angles -75$^{\circ}$ $<$ $\alpha$ $<$+75$^{\circ}$. For wavelengths below 5\,$\mu$m,
the TPM predictions could be off by up to 8\% at specific phase angles
(see inserted plots in Fig.~\ref{fig:BrightnessTempPhase}, bottom).
We also tested trends in the observation-to-model
ratios with the heliocentric distance of the Moon (ranging from 0.985 to 1.018\,au),
the Moon's angular diameter or satellite-Moon distance (ranging from 1756.5 to
1975.7$^{\prime \prime}$, and from 362777.3 to 408034.0\,km, respectively), the 
sub-observer's longitude (351.2 to 6.8$^{\circ}$) and latitude (-5.6 to +6.8$^{\circ}$),
the sub-Solar latitude (-1.6 to +1.5$^{\circ}$), and the calculated
aspect angle (83.5 to 95.6$^{\circ}$), but no obvious correlation has been found.

It is worth noting that our 3\% accuracy in model predictions can be translated
into an approximate error in brightness temperature. At 5\,$\mu$m, a flux density
change of 3\% would translate into a brightness temperature change of 0.8\,K for an
assumed body temperature of 280\,K. At 10\,$\mu$m, the 3\% flux change translates
into 1.6\,K, and at 15\,$\mu$m, the 3\% correspond to 2.4\,K, always referring to a
280\,K object.

\subsection{Interplanetary missions: Thermal IR measurements of the Moon}

\begin{figure}[h!tb]
 \rotatebox{0}{\resizebox{\hsize}{!}{\includegraphics{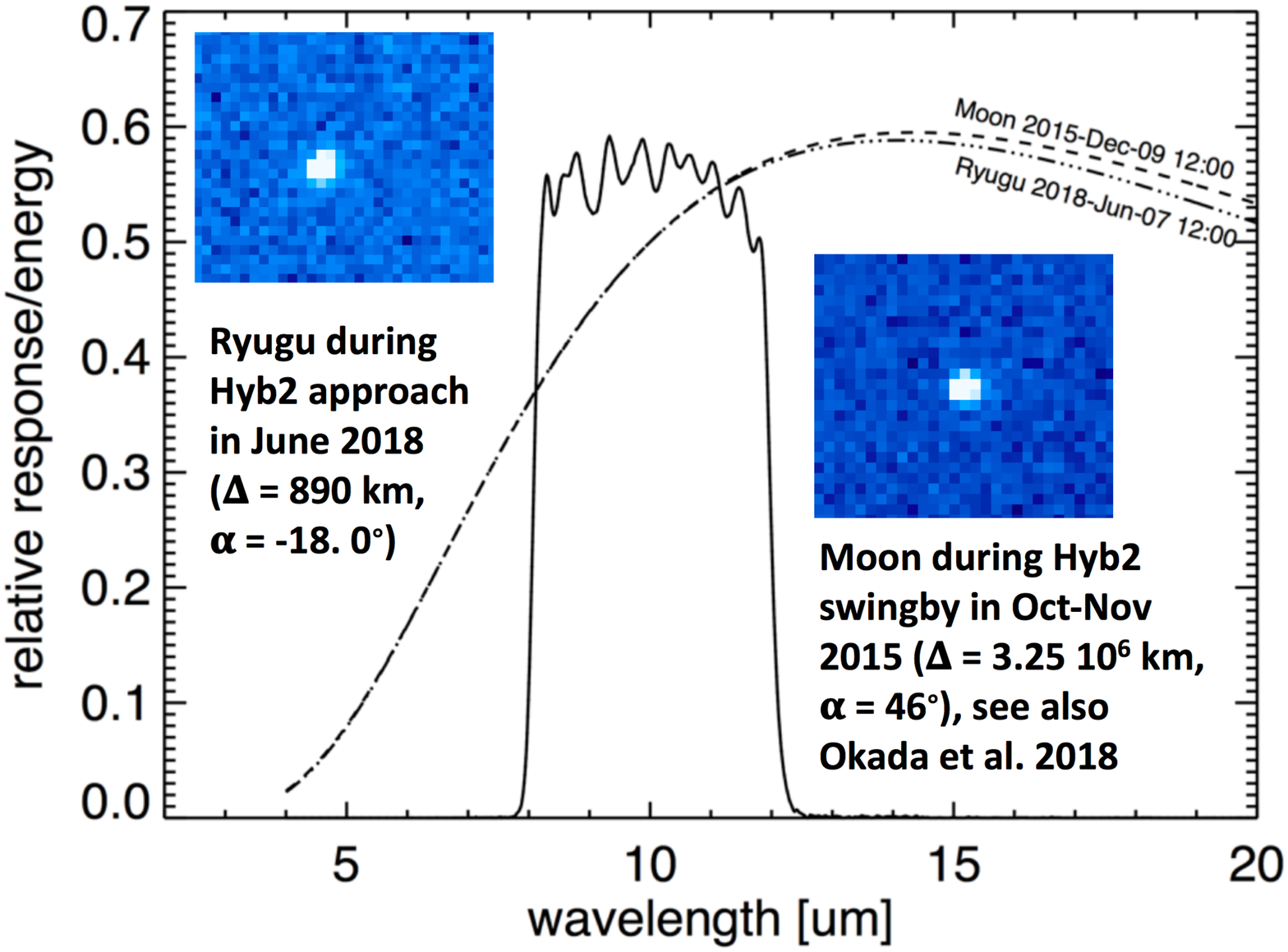}}}
  \caption{Figure shows the TIR passband (in relative detector response
           per energy) in solid line. In addition, the SED shapes of the Moon
           (during the Hayabusa-2 flyby in 2015; dashed line) and Ryugu
           (during close-proximity operations in 2018; dashed-dotted line)
           are shown (arbitrary flux scale). Both SED shapes
           are very similar in the TIR passband. This allows for a direct calibration
           of the TIR Ryugu measurements via (global) lunar model predictions, without
           any color-correction terms.
     \label{fig:Haya2TirMoon}}
\end{figure}

The new thermal model for the global, disk-integrated thermal emission of the Moon
can now be used for calibrating infrared instruments of interplanetary missions. 
The Hayabusa2 mission \citep{Tsuda2016} visited the near-Earth C-type
asteroid 162173~Ryugu and performed a detailed characterization of the
mission target \citep[e.g.,][]{Kitazato2019,Watanabe2019,Sugita2019}.
It revealed the highly porous nature of this primitive asteroid \citep{Okada2020} 
from thermo-graphic images taken by the thermal infrared imager (TIR) \citep{Okada2017}. 
This instrument was calibrated on ground and used for the first scientific measurements
during an Earth-Moon flyby in December 2015 \citep{Okada2018}.
Figure~\ref{fig:Haya2TirMoon} shows the (calculated and arbitrarily scaled)
thermal spectra of the Moon (dashed line) and Ryugu (dashed-dotted line), together with the 
thermal images of both targets as small inlays. Both spectral energy distributions 
are extremely similar in the 5-15\,$\mu$m range despite the very different physical
and thermal properties of these bodies. The TIR passband (shown as solid line)
covers the 8-12\,$\mu$m range. The uncertainties in our global emissivity
solution in the range of the Christiansen feature are very likely not critical
as it is located at the edge of the TIR filter transmission profile.
With the help of our new global lunar model, the
TIR measurements of the Moon (taken at phase angles in the ranges -59 to -55$^{\circ}$
and +45 to +60$^{\circ}$) are currently used to consolidate the TIR in-flight 
detector response and to establish firm correlations between measured signals and
absolute flux densities. The TIR Moon measurements were taken at different distance
from the Moon, covering about three orders of magnitude in count rates. Our TPM 
predictions are taken to characterize the linearity of the detector response
over this wide range, and to understand the size-of-source effect \citep[e.g.,][]{Hill2005,Saunders2013}
which is critical for counts-to-temperature conversion in the TIR calibration process
\citep[see supplementary material in][]{Okada2020}.
More TIR measurements of the Moon have been taken during the return of the Hayabusa-2 sample
container in December 2020, and more are planned in December 2027 and June 2028 during Earth swing-by
maneuvers on its mission extension to asteroids 2001~CC$_{21}$ and 1998~KY$_{26}$.

BepiColombo, an interplanetary mission to Mercury, carries the Mercury Radiometer
and Thermal Infrared Spectrometer (MERTIS).
During an Earth flyby in April 2020 MERTIS collected hyper-spectral 7-14\,$\mu$m
observations of the Moon \citep{Damore2020}. The goal of these measurements was to
test different instrument (acquisition, binning, calibration) modes and to deliver
the first calibrated data acquired in space for the BepiColombo community. Here,
our lessons learnt from the HIRS data and our TPM Moon predictions could be very
useful to reach these goals.

The OSIRIS-REx mission to asteroid Bennu \citep{Lauretta2017} also obtained thermal spectra
of the Moon (at $\alpha$ = 42$^{\circ}$) with the OSIRIS-REx Visible
and InfraRed Spectrometer (OVIRS, 0.4 to 4.3\,$\mu$m) \citep{Simon2019} and the
Thermal Emission Spectrometer (OTES, 5.7 to 100\,$\mu$m) \citep{Christensen2018}
during an Earth flyby in September 2017. The OVIRS full-disk spectra included
the 2.8-$\mu$m hydration band and showed evidence of several other absorption
features. The thermal emission in the range $>$2.8\,$\mu$m had to be subtracted
for the spectral interpretation of the measurements. They modeled the thermal
emission as a single blackbody temperature of 382.5\,K with a wavelength-independent
emissivity of $\epsilon$=0.167. For both, the OVIRS spectra and also the unpublished
OTES data, our full-disk TPM predictions at specific phase angles and wavelengths
would be useful for calibration purposes and the scientific exploitation of the
measurements.
In addition, the MERTIS and the OSIRIS-REx/OTES spectra will be important to
identify any discrepancies between our global emissivity solution at regions not
covered by the HIRS channels.

\subsection{Earth-observing and weather satellites: IR scans of the Moon}

The Moon is also widely used to calibrate space-based Earth observing instruments
\citep[e.g.,][]{Barnes2004,Matthews2008,Burgdorf2019} as a radiative target to
characterize beam properties, to monitor calibration stability, or to inter-compare
detector responses of different instruments or projects. The Clouds and the Earth's
Radiant Energy System (CERES) measures the Earth radiation budget from two polar
orbiting satellites. In addition to high-accuracy broadband measurements of the
scattered and thermal emission, they also perform narrow-band (called WN) 8-12\,$\mu$m
thermal radiance studies. Thanks to regularly obtained Moon scans \citep{Matthews2008}, it was
possible to reduce instrument calibration drifts to 0.3\% per decade and to 
perform inter-instrument comparisons, in addition to studying the instrument's optical degradation.
However, according to \citet{Matthews2008}, the usage of lunar WN data
was suffering from the lack of knowledge about the lunar disk emissivity and
temperature distribution.
Here, our disk-integrated hemispherical emissivity model as well as the calculated
temperature distribution will be very useful
to convert the WN-channel measurements into a broad-band thermal flux and to 
improve the absolute accuracy of WN calibration concepts.
On the other hand, the CERES WN data, typically taken in the phase angle
range 11 to 3$^{\circ}$ before and after opposition, will nicely complement
our lunar TPM testing. Unfortunately, these data were not available for our studies.

The Moon measurements have also been used for the calibration of weather satellites.
\citet{Burgdorf2020} settled a long-standing question about the field of view of the
channels in the HIRS instrument versions 2, 3, and 4. With the contribution of specific Moon intrusions,
they were able to determine the precise pointing direction for each channel
and they found differences of up to 0.031$^{\circ}$ and up to 0.015$^{\circ}$
for the long- and short-wavelength channels, respectively. The photometric agreement
is consistent within about 1\% for most channels and instruments. They also provided
upper limits for the non-linearity effects in the short-wavelength channels.

Other infrared sounders, such as CERES \citep{Daniels2014} or
IASI\footnote{Infrared Atmospheric Sounding Interferometer;
\url{https://www.mdpi.com/2072-4292/12/9/1488}},
have extensively observed the Moon, and future instrument generations, such as the
IASI-New Generation or the Meteorological Imager on Metop Second Generation,
will also see the Moon. This makes our model an ideal tool for cross-calibration
exercises. It can also provide the grounds for monitoring instrument
or detector aging and degradation effects.

Based on the available HIRS Moon detections, we intercompared the calculated ratios
for a given channel (or a small subset of channels) per instrument (HIRS/2, HIRS/3,
and HIRS/4). We noticed that the HIRS/3 ratios are on average about 5\% lower than
the HIRS/2 or HIRS/4 ratios. But the statistics for HIRS/3 are poor and some of the
HIRS/3 data are flagged with the Moon potentially touching the edge the FOV. In
addition, all the HIRS/2 data were taken at negative phase angles (-71$^{\circ}$ to
-34$^{\circ}$), while all the HIRS/3 measurements are in the positive phase angle
range (+23$^{\circ}$ to +73$^{\circ}$). Only the HIRS/4 data cover the full range
of angles. However, with the availability of more Moon intrusions, this would be the
right path to inter-compare the absolute calibration of the instruments, find detector
aging effects, or test possible small deviations in the beam size.

\subsection{Thermal modeling of asteroids}

With the availability of large sets of thermal asteroid observations taken at short
wavelengths, there is a need to attain a better understanding of the applied radiometric
techniques. The HIRS data (Tables~\ref{tbl:hirs_obs_app_long} and \ref{tbl:hirs_obs_app_short})
are excellent for
testing and verifying thermal models and for characterizing possible error of the derived
parameters. The analysis of the HIRS Moon data reveal the critical concepts and
shortcomings for the application to asteroid observations at short thermal wavelengths:

First, in cases where asteroids show regions with different albedos, the
standard radiometric size-albedo solutions (using only short-wavelength
thermal data) will be biased towards the lowest albedo values.
Hence, albedo variations can bias the size determination to larger
sizes\footnote{A low-albedo small object can have the
same thermal flux as a higher-albedo and larger body.}.

Second, surface roughness - at scales that are larger than the thermal skin
depth and smaller than the resolution of the global shape model - is
crucial for the interpretation of observations taken well below the thermal
emission peak and at large phase angles (mainly relevant for near-Earth objects;
see also Fig.~\ref{fig:RoughnessTest}). Uncertainty in the emissivity properties
leads to unrealistic roughness values (see the case of Ryugu below).

Third, unknown hemispherical emissivity properties introduce substantial errors
in the radiometric determination of size-albedo solutions when using only
short-wavelength observations. In addition, in cases where thermal data cover a
wide range in wavelengths that also includes mid-IR data, radiometric studies
can lead to erroneous object properties when a constant spectral emissivity is assumed.

Next, laboratory spectral emissivity studies of typical surface materials might
not have a one-to-one correspondence to full-disk spectra. There are
also strong vacuum effects on emissivity spectra.

And last, for near-Earth objects, reflected sunlight contributions must be
accounted for before interpreting thermal measurements below $\approx$5\,$\mu$m.

\subsubsection{NEATM Applications}

The near-Earth asteroid thermal model (NEATM; \citet{Harris1998}) is widely used for
radiometric studies of asteroids \citep[see][for a review]{Harris2002}.
It assumes a non-rotating spherical object with a non-conducting, perfectly smooth
surface with constant emissivity and albedo. To compensate for these simplifying
assumptions, a single parameter $\eta$, called the beaming parameter, can either be
fitted to the thermal emission spectrum or used with a default value when
multi-wavelength data are not available. A value of 1.0 has been widely used
for main-belt asteroids observed at phase angles well below 45\,$^{\circ}$ and
1.5 for larger phase angles typical of near-Earth asteroids, but other values have
been applied in the past \citep[e.g.,][]{Delbo2007, Wolters2009, Mueller2011, Mainzer2016, Ali-Lagoa2018, Mommert2018, Mueller2020}.

In recent years, IR sky surveys such as WISE or dedicated projects with Spitzer-IRAC
produced a wealth of data in the 3.5 to 5\,$\mu$m range.
The NEATM applications at short thermal wavelengths and in the presence of reflected
sunlight are addressed in different works related to NEOWISE data
or Spitzer-IRAC observations \citep[e.g.,][and references therein]{Mainzer2015},
but the correct application techniques and potential errors are under discussion
\citep{Mommert2018, Myhrvold2018, Wright2018, Masiero2021}.
Here, the HIRS channels 13-19 provide a very useful means of investigating emissivity
properties that are different from the standard $\epsilon$-solutions with constant
values of 0.9 or 0.95.

For the HIRS data of the Moon, we find $\eta$-values of 0.9 for the smallest phase angle of
15.6$^{\circ}$ up to values of 1.2 at the largest phase angles $>$70$^{\circ}$.
It is important to note that for a given multi-wavelength observation, the
fitted $\eta$ can vary significantly, depending on the given subset of channels
used in the analysis.
At short wavelengths (without ch18 and ch19, where reflected sunlight is
contributing), we see $\approx$5-10\% higher $\eta$ values, up to about 1.25
at short wavelengths and large phase angles.
In the long-wavelength channels we find systematically lower $\eta$-values,
down to about 0.8 at $\lambda$ $>$ 12\,$\mu$m and small phase angles.
Also, we found that fitting $\eta$ to data below 5\,$\mu$m is very uncertain
and leads to huge errors in size-albedo solutions.
If we take the full-wavelength-fitted best $\eta$-values (between 0.9
and 1.2, depending on the phase angle: see above) and apply the NEATM to the HIRS data,
we can reproduce the Moon's size to within 5-10\% (and the albedo to within
10-20\%) at all wavelengths beyond about 4\,$\mu$m, which is consistent with
the error bars expected for the model when applied to asteroids (see e.g., the
discussion in \citealt{Ali-Lagoa2018}). This error range is only applicable
in cases where the available thermal data allow for a high-quality $\eta$ fit
and where the thermal parameter \citep{Spencer1989} is small. For much faster
rotating objects, high thermal inertia surfaces, or complex shapes, the
NEATM size-albedo solutions are naturally more uncertain.

\subsubsection{TPM Applications}

The TPM techniques discussed above are typically applied to asteroids with known spin (and possibly also shape) 
information. The goal is to determine the object's true size and albedo and to constrain its
thermal inertia and surface roughness. In some cases, the TPM analysis also helps to 
improve the quality of the spin and shape solutions \citep[e.g.,][]{Mueller2017, Mueller2018}.
But the success of these methods 
strongly depends on the availability and quality of thermal measurements.
Based on our HIRS data, we can perform a radiometric study of the Moon by simply using
its known spin and shape. But we also consider how well such a procedure reproduce the Moon's size without
having any information about its hemispherical spectral emissivity. We consider what types of constraints
on thermal inertia and surface roughness we may obtain by using a typical approach used for asteroids.

Running a standard TPM radiometric analysis on the HIRS data produces
results with severe implications for asteroid studies. We leave size,
albedo, thermal inertia, and surface roughness as free parameters. We
take a spherical shape with the Moon's true spin properties (spin-axis
orientation and synodic rotation period) and assume a constant
spectral emissivity of $\epsilon$ = 0.9 (or 0.95), as is frequently
used in asteroid studies.

The first finding, which was already expected from the Moon's very slow rotation,
is that the thermal inertia cannot be determined from our data. Low
$\Gamma$-values of 10 SI-units produce the same size-albedo solution
as $\Gamma$-values well above 100 SI-units. However, the HIRS data
include a strong signature of surface roughness. At short wavelengths
(and to a lesser extent, also at long wavelengths), an intermediate level
of surface roughness is needed (close to rms of 30$^{\circ}$) to
produce consistent size-albedo solutions (with acceptable $\chi^2$ values)
over all phase angles.
With such an intermediate level of surface roughness, the LW data
($\lambda$-range from 6.5 to 15\,$\mu$m) directly produce
a radiometric size which is within 10\,km of the Moon's true diameter
(3488 vs.\ 3474.8\,km). However, at short-wavelength data, the
constant-$\epsilon$ assumption leads to a size which is about 15\% lower,
that is, well below 3000\,km,  even after
cleaning the SW data from the channels below 4\,$\mu$m and the extreme
phase angles ($<$-70$^{\circ}$ and $>$+70$^{\circ}$). 
At short wavelengths, we need lower roughness (down to a smooth surface) to match the
radiometric size to the Moon's true size. But this violates the
constraints from the phase curve. A smooth surface is incompatible
with the measured flux change with phase angle. The solution to the
problem is clearly related to the Moon's lower spectral emissivity
at short wavelengths. 
This is an important aspect when trying to apply radiometric
techniques to short-wavelength, often single-epoch, asteroid data.
\citet{Mueller2017} analyses a collection of pre-mission thermal
measurements of 162173 Ryugu, the Hayabusa-2 target asteroid. The
data set was dominated by short-wavelength Spitzer-IRAC data at
3.55 and 4.49\,$\mu$m. And as a result, the best-fit radiometric
solution pointed towards a smooth surface. This setting was
required to bring the short- and long-wavelength data into
agreement and to obtain an acceptable $\chi^2$ size-albedo
solution\footnote{\citet{Mueller2017} derived an effective diameter
of 850 to 880\,m, a geometric albedo of 0.044 to 0.050, and a thermal
inertia $\Gamma$ in the range 150 to 300\,J m$^{-2}$s$^{-0.5}$K$^{-1}$
which is in excellent agreement with Ryugu's true properties
from in-situ studies.}. Recently, the Hayabusa-2 data revealed a surface
roughness rms of 47$^{\circ}$ $\pm$ 5$^{\circ}$ \citep{Shimaki2020}.
This high level of surface roughness (compared to the 32$^{\circ}$
for the Moon) is in contradiction to the smooth surface predicted
by \citet{Mueller2017} and points toward shortcomings in their
radiometric study. We repeated the analysis of the pre-mission IR
measurements of Ryugu (the 24 best-quality measurements from
\citet{Mueller2017}), but now using the in-situ size, shape, spin
\citep{Watanabe2019}, and surface roughness properties
(47$^{\circ}$ $\pm$ 5$^{\circ}$; \citet{Shimaki2020}). 

In a first approach, we used standard
constant spectral emissivity of 0.9, then, in a second step, we applied
the HIRS-based lunar spectral emissivity, always with the goal of constraining
the object's thermal inertia via $\chi^2$ minimization techniques
\citep[e.g.,][]{Ali-Lagoa2020}.
On the one hand, the default $\epsilon$=0.9 assumption
leads to a thermal inertia above 1000\,J m$^{-2}$s$^{-0.5}$K$^{-1}$ (for
acceptable reduced $\chi^2$ values below 2).
On the other hand, with the lunar global emissivity model we found an acceptable
minimum $\chi^2$ close to 1.0. The corresponding thermal inertia is
between 150 and 400\,J m$^{-2}$s$^{-0.5}$K$^{-1}$, very
close to the published values of about 300\,J m$^{-2}$s$^{-0.5}$K$^{-1}$
\citep{Okada2020} and 225 $\pm$ 45\,J m$^{-2}$s$^{-0.5}$K$^{-1}$ \citep{Shimaki2020}.
This example illustrates the importance of realistic spectral emissivity in
the context of radiometric studies based on short-wavelength ($<$10\,$\mu$m)
thermal measurements.

\section{Conclusions}
\label{sec:con}

With our collection of NOAA- and MetOp-HIRS Moon detections,
we are able to provide a unique thermal dataset for characterizing the
global thermal properties of the Moon and for testing and benchmarking
asteroid thermal models. 
The HIRS full-disk lunar fluxes cover a wavelength range from
3.75 to 15.0\,$\mu$m and a phase angle range from -73.1$^{\circ}$
(waxing Moon) to +73.8$^{\circ}$ (waning Moon), taken between
1989 and 2019. The measurements are absolutely calibrated
with an estimated uncertainty of 3\% or less, except in the
few cases where the Moon was possibly touching the edge of
a given channel FOV.
Since these measurements are dominated by thermal emission of
the sunlit regions, they have no or very little information
about the Moon's global thermal inertia. However, they contain
a strong signature of the global surface roughness properties.
This is seen in characteristic flux changes with phase angle
which point to a roughness rms of surface slopes close to 32$^{\circ}$,
confirming the findings of \citet[and references therein]{Rozitis2011}.

The correct roughness description allows us to combine the HIRS data
taken over this very wide phase angle range from almost first quarter
(waxing) to third quarter (waning). In a second step, it was then
possible to examine the Moon's hemispherical spectral emissivity.
Emissivity values determined from the HIRS data
lead to a partial match to the mean characteristics of lunar mare
and highlands at very short and very long HIRS wavelengths.
The global emissivity solution shows values as low
as 0.69 at 4.5\,$\mu$m and then slowly increasing to values close
to 1.0 at 12\,$\mu$m.
In the intermediate wavelength range, at 6.5\,$\mu$m (channel 12), 
7.3\,$\mu$m (channel 11), and 9.7\,$\mu$m (channel 09), the
HIRS-based emissivity is about 5-10\% lower than the values derived
from the Apollo samples of lunar mare, highlands and transition
regions. In the region of the Christiansen feature
(between channel 11 and channel 09) significantly higher emissivity
values are expected, but this range is not covered by the HIRS channels.

The combined HIRS data helped us to establish a global
lunar TPM. This model solution allows us to predict the
Moon's full-disk emission for a wide range of phase angles
and wavelengths, with the exception of the region 
covered by the Christiansen feature.
Our model predictions reach an absolute
accuracy of better than about 5\% in the mid-IR regime
(and better than about 10\% at the shortest wavelengths
below 5\,$\mu$m). We still see an asymmetric emission before
and after opposition 
and our model starts to deviate a few percent from observations at 
the extreme phase angles ($|\alpha|>75^\circ$) based on the given
wavelength range and whether the phase is waxing or 
waning (see Sec.~\ref{sec:phasecurves}). However, for highly
accurate model applications, it is possible to correct these
small offsets by manual corrections based on our linear fits.

The applications of the HIRS data are manifold.
Thermal-IR instrumentation 
on weather satellites and interplanetary missions are using
the Moon as in-flight calibrator to improve the knowledge on
beam characteristics or to verify detector response and
linearity properties.
What is of particular interest is the potential for inter-calibration and
checking the photometric stability of instrumentation on weather
satellites. The Moon appeared in the DSV of
HIRS/2 as early as 1983. The radiances measured back then can be
compared immediately with those obtained in 2020 by HIRS/4 on Metop
satellites, when a model can provide the exact dependence of the
radiance from the phase angle of the Moon and its distance to the
Sun. Such a model offers therefore the intriguing  possibility of
reliable time series of essential climate variables over almost four
decades. With  respect to HIRS, here we think, in particular, of the
upper tropospheric humidity, whose response to CO$_{2}$ forcing plays
an important role in climate models.

However, our model will not only be important for re-calibrating existing
infrared satellite observation records for climate research applications,
but also for the upcoming climate satellite mission
FORUM\footnote{Far-infrared Outgoing Radiation Understanding
and Monitoring} \citep{Palchetti2020}
and other climate-oriented infrared Earth observing missions, for which
absolute radiometric accuracy and stability are crucial.
The HIRS data are also very useful
for testing and benchmarking asteroid thermal models. Here, the
influence of surface roughness and spectral hemispherical
emissivity are crucial. In the particular context
of short-wavelength near-Earth asteroid measurements,
the quality of the derived radiometric properties can
be improved. In this work, we demonstrate the impact of spectral
emissivity for the determination of the thermal inertia and
surface roughness for Ryugu.

\begin{acknowledgements}
TM and VAL have received funding from the European
Union's Horizon 2020 Research and Innovation Programme, under Grant Agreement
no.\ 687378, as part of the project "Small Bodies Near and Far" (SBNAF).
MB has received funding from the Deutsche Forschungsgemeinschaft, project number
421761264. With this work we contribute to the Cluster of Excellence "CLICCS—Climate,
Climatic Change, and Society" funded by the Deutsche Forschungsgemeinschaft DFG
(EXC 2037, Project Number 390683824), and to the Center for Earth System Research
and Sustainability (CEN) of Universit{\"a}t Hamburg.
TM also acknowledges the financial support and hospitality of the International
Space Science Institute (ISSI) in Bern, Switzerland, which initiated this work
by sponsoring and hosting our "ISSI-team" meetings during 2011 to 2014.
The level 1b files of HIRS were read and processed using Typhon \citep{Lemke2020}.
\end{acknowledgements}

\bibliographystyle{aa}  
\bibliography{AsteroidsGeneral}

\clearpage
\newpage
\begin{appendix}

\section{Thermal-IR HIRS measurements of the Moon}
\label{app:hirs_obs}

We list the extracted and derived HIRS Moon values in the following tables.
Tables~\ref{tbl:hirs_obs_app_tbl1}, \ref{tbl:hirs_obs_app_tbl2}, and \ref{tbl:hirs_obs_app_tbl3}
list the calibrated radiances (Rad) and brightness temperatures (T$_b$) of the Moon for
the long-wavelength channels 1-6, 7-12 and the short-wavelength channels 13-19, respectively.
Due to the unknown contribution of reflected sunlight in channels 17 and 18, there were no brightness
temperatures calculated in these two channels.

The disk-integrated flux densities (in Jansky) were calculated from the radiance values and the
apparent cross-section of the Moon at the time of the observation and as seen from the
particular satellite. Tables~\ref{tbl:hirs_obs_app_long} and \ref{tbl:hirs_obs_app_short}
contain all crucial values for the long-wavelength channels 1-12 and the short-wavelength
channels 13-19, respectively.

The "observation epoch" data and time entries are taken from the coordinate "time" in
the level 1b data file. We added 3.2\,sec to this value, to give the time of the
middle of the scan instead of its start. The HIRS scans were manually inspected to
identify the range of scan positions, where the signal in the DSV was constant.
This is the case if and only if the Moon is fully included in the FOV or not present
at all. A few cases where the visual inspection left some doubt are labeled with
footnotes in Tbl.~\ref{tbl:hirs_events}, that is,\ the measured flux is only a lower limit.
These limits, however, are always rather close to the real value. Details on all
detections can all be provided
on request, including the range of scan positions, mean counts from the DSV with the Moon
included (as well as 256\,sec before and after during the calibration configurations),
the standard deviation of the counts, the mean temperatures of the internal warm calibrator
target from level 1b data (five temperatures for HIRS/2 and HIRS/3, and six temperatures
for HIRS/4), and the fitted blackbody radiances and slopes.

The radiance values of the Moon were calculated with a two-point calibration
using "slope" and Moon-DSV differences, then divided by the fraction of the FOV
filled by the Moon and the beam efficiency (\citet{Koenig1980},
and User's Guide\footnote{\url{http://rain.atmos.colostate.edu/XCAL/docs/amsub/NOAA KLM Users Guide.pdf}}).
Its $\sigma$ was calculated with the error propagation from the counts, assuming that
the error of the BB Temp is zero. The "brightness temperatures" are derived from the radiance
via the inverse Planck function at the specified wavelength\footnote{The central wavelength of
the channels are given on the web page
\url{https://www.nwpsaf.eu/site/software/rttov/download/coefficients/spectral-response-functions/}.}.
We used the shifted relative spectral response functions.
The FOV diameters, assumed in the calculation of radiance, were taken from the OSCAR
web page and confirmed via the JPL/Horizons system (see also Section~\ref{sec:hirs}).

\subsection{Calibrated Radiances and brightness temperatures}
\subsection{Moon full-disk flux densities}


\begin{landscape}
  \begin{table}[h!tb]
    \begin{center}
{\tiny
\caption{Calibrated Radiances (Rad) and brightness temperatures (T$_b$) of the Moon for the long-wavelength channels 1-6.
             \label{tbl:hirs_obs_app_tbl1}}
    \begin{tabular}{l|rrrrrrrrrrrr}
      \hline
      \hline
      \noalign{\smallskip}
Satellite/Instrument & \multicolumn{2}{c}{$\lambda_{ch01}$} & \multicolumn{2}{c}{$\lambda_{ch02}$} & \multicolumn{2}{c}{$\lambda_{ch03}$} & \multicolumn{2}{c}{$\lambda_{ch04}$} & \multicolumn{2}{c}{$\lambda_{ch05}$} & \multicolumn{2}{c}{$\lambda_{ch06}$} \\
Observation Epoch    & Rad [MJy/sr] & T$_b$ [K]        & Rad [MJy/sr] & T$_b$ [K]        & Rad [MJy/sr] & T$_b$ [K]        & Rad [MJy/sr] & T$_b$ [K]        & Rad [MJy/sr] & T$_b$ [K]        & Rad [MJy/sr] & T$_b$ [K] \\
      \noalign{\smallskip}
      \hline
      \noalign{\smallskip}

NOAA-11/HIRS2    & \multicolumn{2}{c}{14.852} & \multicolumn{2}{c}{14.731} & \multicolumn{2}{c}{14.488} & \multicolumn{2}{c}{14.183} & \multicolumn{2}{c}{13.909} & \multicolumn{2}{c}{13.658} \\
1993-11-25 17:39:34 & 7.0029E+08 & 333.1 &  7.1973E+08 &  336.2 &  7.1915E+08 &  336.3 &  7.1057E+08 &  335.3 &  6.9957E+08 & 334.1 &  6.9930E+08 & 334.6 \\
1992-02-15 19:54:15 & 7.4778E+08 & 340.4 &  7.7230E+08 &  344.1 &  7.7500E+08 &  344.5 &  7.6403E+08 &  343.1 &  7.5399E+08 & 342.0 &  7.5797E+08 & 343.0 \\
1990-11-27 05:04:49 & 4.0605E+08 & 282.5 &  4.1279E+08 &  284.1 &  4.1873E+08 &  286.0 &  4.0885E+08 &  285.2 &  4.0393E+08 & 285.3 &  4.0205E+08 & 286.0 \\
1989-03-17 03:31:08 & 5.0036E+08 & 300.1 &  5.2402E+08 &  304.5 &  5.3369E+08 &  306.7 &  5.2788E+08 &  306.5 &  5.2205E+08 & 306.3 &  5.2341E+08 & 307.4 \\

\noalign{\smallskip}
    \hline
\noalign{\smallskip}

NOAA-14/HIRS2    & \multicolumn{2}{c}{14.955} & \multicolumn{2}{c}{14.701} & \multicolumn{2}{c}{14.487} & \multicolumn{2}{c}{14.171} & \multicolumn{2}{c}{13.931} & \multicolumn{2}{c}{13.656} \\
2000-04-14 10:50:44 & 5.5529E+08 & 309.4 &  5.6072E+08  &  310.9 &  5.6217E+08 &  311.5 &  5.6149E+08 &  312.1 &  5.6476E+08 & 313.3 &  5.6277E+08 & 313.8 \\
1997-12-08 21:52:12 & 4.1319E+08 & 283.6 &  4.1789E+08  &  285.2 &  4.2159E+08 &  286.6 &  4.1879E+08 &  287.1 &  4.2175E+08 & 288.6 &  4.2141E+08 & 289.6 \\
1997-06-15 05:05:20 & 3.7208E+08 & 275.5 &  3.9263E+08  &  280.2 &  3.9936E+08 &  282.3 &  3.9579E+08 &  282.7 &  3.9949E+08 & 284.4 &  4.0078E+08 & 285.8 \\
1996-10-23 00:30:08 & 6.3513E+08 & 322.8 &  6.2982E+08  &  322.2 &  6.3101E+08 &  322.7 &  6.3170E+08 &  323.4 &  6.3456E+08 & 324.3 &  6.3426E+08 & 324.9 \\
1996-05-28 20:58:58 & 5.7402E+08 & 312.6 &  5.5256E+08  &  309.5 &  5.5147E+08 &  309.7 &  5.5032E+08 &  310.3 &  5.5267E+08 & 311.3 &  5.5170E+08 & 312.0 \\
1995-12-03 07:13:24 & 6.9262E+08 & 331.9 &  6.9824E+08  &  332.9 &  7.0785E+08 &  334.5 &  7.0848E+08 &  335.0 &  7.1355E+08 & 336.1 &  7.1490E+08 & 336.8 \\

\noalign{\smallskip}
    \hline
\noalign{\smallskip}

NOAA-15/HIRS3    & \multicolumn{2}{c}{14.976} & \multicolumn{2}{c}{14.743} & \multicolumn{2}{c}{14.508} & \multicolumn{2}{c}{14.226} & \multicolumn{2}{c}{13.963} & \multicolumn{2}{c}{13.666} \\
2019-12-14 09:47:24 & NaN         & NaN   &  NaN         &  NaN   &  NaN        &  NaN   &  NaN        &  NaN   &  NaN        & NaN   &  NaN        & NaN   \\
2018-12-24 11:44:46 & 7.9469E+08  & 347.5 &  8.0152E+08  &  348.4 &  8.0158E+08 &  348.4 &  8.0388E+08 &  348.8 &  8.0714E+08 & 349.5 &  8.0785E+08 & 349.9 \\

\noalign{\smallskip}
    \hline
\noalign{\smallskip}

NOAA-17/HIRS3    & \multicolumn{2}{c}{14.966} & \multicolumn{2}{c}{14.669} & \multicolumn{2}{c}{14.474} & \multicolumn{2}{c}{14.219} & \multicolumn{2}{c}{13.958} & \multicolumn{2}{c}{13.664} \\
2003-01-24 01:20:02 & 3.4799E+08 &  270.4 &  3.5665E+08 &  273.2 &  3.5572E+08 &  273.6 &  3.6154E+08 &  275.8 &  3.5885E+08 & 276.3 &  3.5869E+08 & 277.6 \\
2002-09-26 07:01:22 & 4.9181E+08 &  298.0 &  5.1313E+08 &  302.9 &  5.2349E+08 &  304.9 &  5.3267E+08 &  307.2 &  5.3600E+08 & 308.5 &  5.4329E+08 & 310.6 \\

\noalign{\smallskip}
    \hline
\noalign{\smallskip}

NOAA-18/HIRS4    & \multicolumn{2}{c}{14.980} & \multicolumn{2}{c}{14.697} & \multicolumn{2}{c}{14.513} & \multicolumn{2}{c}{14.223} & \multicolumn{2}{c}{14.006} & \multicolumn{2}{c}{13.670} \\
2018-11-26 17:59:54 & 5.9202E+08 &  315.7 &  6.1874E+08  &  320.4 &  6.2177E+08 &  321.2 &  6.2056E+08 &  321.5 &  6.1958E+08 & 321.8 &  6.1922E+08 & 322.6 \\
2018-04-02 09:27:54 & 7.8316E+08 &  345.7 &  7.9485E+08  &  347.4 &  7.9684E+08 &  347.7 &  7.9588E+08 &  347.7 &  7.9542E+08 & 347.8 &  7.9757E+08 & 348.5 \\
2011-03-14 10:32:05 & 3.6979E+08 &  274.8 &  3.6765E+08  &  275.3 &  3.6765E+08 &  275.9 &  3.6543E+08 &  276.5 &  3.6336E+08 & 277.0 &  3.6302E+08 & 278.4 \\

\noalign{\smallskip}
    \hline
\noalign{\smallskip}


NOAA-19/HIRS4    & \multicolumn{2}{c}{14.953} & \multicolumn{2}{c}{14.685} & \multicolumn{2}{c}{14.526} & \multicolumn{2}{c}{14.232} & \multicolumn{2}{c}{13.973} & \multicolumn{2}{c}{13.635} \\
2017-12-02 12:50:03 & 8.7728E+08 &  359.4 &  8.7793E+08  &   359.3 &  8.7801E+08 &  359.2 &  8.7378E+08 &  358.5 &  8.7204E+08 & 358.4 &  8.7601E+08 & 359.1 \\
2012-03-04 05:07:04 & NaN        &  NaN   &  NaN         &   NaN   &  NaN        &  NaN   &  NaN        &  NaN   &  NaN        & NaN   &  NaN        & NaN   \\

\noalign{\smallskip}
    \hline
\noalign{\smallskip}


MetOp-A/HIRS4    & \multicolumn{2}{c}{14.955} & \multicolumn{2}{c}{14.719} & \multicolumn{2}{c}{14.503} & \multicolumn{2}{c}{14.248} & \multicolumn{2}{c}{13.955} & \multicolumn{2}{c}{13.667} \\
2011-11-16 23:42:26 & 3.8160E+08 & 277.3 &  3.8498E+08 & 278.7 & 3.8664E+08 & 279.7 & 3.8605E+08 & 280.5 & 3.8720E+08 & 281.9 & 3.8538E+08 & 282.8 \\
\noalign{\smallskip}
    \hline
\noalign{\smallskip}


MetOp-B/HIRS4    & \multicolumn{2}{c}{14.964} & \multicolumn{2}{c}{14.688} & \multicolumn{2}{c}{14.474} & \multicolumn{2}{c}{14.248} & \multicolumn{2}{c}{13.996} & \multicolumn{2}{c}{13.671} \\
2019-07-21 07:57:26 & 5.9845E+08 &  319.0 &  6.0140E+08 &  319.9 &  5.9840E+08 &  319.8 &  5.9717E+08 &  320.0 &  5.9315E+08 & 319.9 &  5.9208E+08 & 320.6 \\
2015-02-07 05:12:40 & 7.7175E+08 &  344.0 &  7.6936E+08 &  343.6 &  7.6997E+08 &  343.8 &  7.6840E+08 &  343.7 &  7.6449E+08 & 343.4 &  7.6443E+08 & 343.8 \\

      \noalign{\smallskip}
      \hline
    \end{tabular}
}
    \end{center}
  \end{table}
\end{landscape}

\begin{landscape}
  \begin{table}[h!tb]
    \begin{center}
{\tiny
    \caption{Calibrated Radiances and brightness temperatures of the Moon for the long-wavelength channels 7-12.
             \label{tbl:hirs_obs_app_tbl2}}
    \begin{tabular}{l|rrrrrrrrrrrr}
      \hline
      \hline
      \noalign{\smallskip}
Satellite/Instrument & \multicolumn{2}{c}{$\lambda_{ch07}$} & \multicolumn{2}{c}{$\lambda_{ch08}$} & \multicolumn{2}{c}{$\lambda_{ch09}$} & \multicolumn{2}{c}{$\lambda_{ch10}$} & \multicolumn{2}{c}{$\lambda_{ch11}$} & \multicolumn{2}{c}{$\lambda_{ch12}$ [$\mu$m]} \\
Observation Epoch    & Rad [MJy/sr] & T$_b$ [K]        & Rad [MJy/sr] & T$_b$ [K]        & Rad [MJy/sr] & T$_b$ [K]        & Rad [MJy/sr] & T$_b$ [K]        & Rad [MJy/sr] & T$_b$ [K]        & Rad [MJy/sr] & T$_b$ [K] \\

      \noalign{\smallskip}
      \hline
      \noalign{\smallskip}

NOAA-11/HIRS2    & \multicolumn{2}{c}{13.340} & \multicolumn{2}{c}{11.088} &  \multicolumn{2}{c}{9.733} & \multicolumn{2}{c}{12.565} &  \multicolumn{2}{c}{7.362} &  \multicolumn{2}{c}{6.738} \\
1993-11-25 17:39:34 & 6.9877E+08 & 335.3 & 6.3100E+08 & 336.7 &  5.6741E+08 & 340.4 &  6.8019E+08 & 335.2 &  3.2155E+08 &  340.5 &  2.4419E+08 &  340.1 \\
1992-02-15 19:54:15 & 7.5477E+08 & 343.2 & 6.8889E+08 & 344.3 &  6.2065E+08 & 347.5 &  7.3917E+08 & 343.3 &  3.5250E+08 &  346.1 &  2.6775E+08 &  345.2 \\
1990-11-27 05:04:49 & 4.0074E+08 & 287.2 & 3.5113E+08 & 292.9 &  3.0803E+08 & 298.8 &  3.8745E+08 & 288.8 &  1.6844E+08 &  306.1 &  1.2428E+08 &  307.1 \\
1989-03-17 03:31:08 & 5.1980E+08 & 307.9 & 4.6747E+08 & 312.8 &  4.1497E+08 & 317.7 &  5.0691E+08 & 309.3 &  2.3004E+08 &  321.8 &  1.7181E+08 &  322.1 \\

\noalign{\smallskip}
    \hline
\noalign{\smallskip}

NOAA-14/HIRS2    & \multicolumn{2}{c}{13.317} & \multicolumn{2}{c}{11.119} &  \multicolumn{2}{c}{9.725} & \multicolumn{2}{c}{12.508} &  \multicolumn{2}{c}{7.398} &  \multicolumn{2}{c}{6.733} \\
2000-04-14 10:50:44 & 5.5863E+08 & 314.3 & 4.9360E+08 & 316.6 &  4.2576E+08 & 319.6 &  5.4063E+08 & 315.0 &  2.4988E+08 &  325.4 &  1.8643E+08 &  326.2 \\
1997-12-08 21:52:12 & 4.1836E+08 & 290.5 & 3.6626E+08 & 295.4 &  3.1360E+08 & 299.9 &  4.0375E+08 & 292.1 &  1.8165E+08 &  309.0 &  1.3416E+08 &  310.7 \\
1997-06-15 05:05:20 & 3.9883E+08 & 286.9 & 3.5284E+08 & 292.9 &  3.0265E+08 & 297.8 &  3.8658E+08 & 289.0 &  1.7057E+08 &  305.9 &  1.2469E+08 &  307.4 \\
1996-10-23 00:30:08 & 6.2580E+08 & 324.6 & 5.6161E+08 & 326.7 &  4.9236E+08 & 329.8 &  6.0748E+08 & 325.1 &  2.8655E+08 &  333.1 &  2.1521E+08 &  333.5 \\
1996-05-28 20:58:58 & 5.4208E+08 & 311.7 & 4.8369E+08 & 315.1 &  4.2213E+08 & 319.0 &  5.2410E+08 & 312.4 &  2.4065E+08 &  323.4 &  1.7826E+08 &  324.0 \\
1995-12-03 07:13:24 & 7.1163E+08 & 337.2 & 6.4413E+08 & 338.2 &  5.6498E+08 & 340.1 &  6.9442E+08 & 337.4 &  3.3636E+08 &  342.4 &  2.5264E+08 &  342.1 \\

\noalign{\smallskip}
    \hline
\noalign{\smallskip}

NOAA-15/HIRS3    & \multicolumn{2}{c}{13.357} & \multicolumn{2}{c}{11.151} &  \multicolumn{2}{c}{9.705} & \multicolumn{2}{c}{12.444} &  \multicolumn{2}{c}{7.303} &  \multicolumn{2}{c}{6.534} \\
2019-12-14 09:47:24 & NaN        & NaN   &  NaN       &  NaN  &   NaN       &  NaN  &   NaN       &  NaN  &   NaN       &   NaN  &   NaN       &   NaN  \\
2018-12-24 11:44:46 & 8.0037E+08 & 349.4 & 7.4721E+08 & 351.4 &  6.1546E+08 & 347.1 &  7.9463E+08 & 351.0 &  3.6596E+08 &  349.7 &  2.6748E+08 &  350.7 \\

\noalign{\smallskip}
    \hline
\noalign{\smallskip}

NOAA-17/HIRS3    & \multicolumn{2}{c}{13.360} & \multicolumn{2}{c}{11.123} &  \multicolumn{2}{c}{9.722} & \multicolumn{2}{c}{12.430} &  \multicolumn{2}{c}{7.320} &  \multicolumn{2}{c}{6.547} \\
2003-01-24 01:20:02 & 3.5210E+08 & 277.7 & 3.0859E+08 & 284.3 &  2.4552E+08 & 285.9 &  3.4399E+08 & 281.4 &  1.3770E+08 &  297.7 &  9.5449E+07 &  301.0 \\
2002-09-26 07:01:22 & 5.4271E+08 & 311.6 & 5.0553E+08 & 318.4 &  4.2115E+08 & 318.9 &  5.3860E+08 & 315.0 &  2.3880E+08 &  324.7 &  1.6790E+08 &  326.1 \\

\noalign{\smallskip}
    \hline
\noalign{\smallskip}

NOAA-18/HIRS4    & \multicolumn{2}{c}{13.340} & \multicolumn{2}{c}{11.117} &  \multicolumn{2}{c}{9.720} & \multicolumn{2}{c}{12.501} &  \multicolumn{2}{c}{7.314} &  \multicolumn{2}{c}{6.513} \\
2018-11-26 17:59:54 & 6.2009E+08 &  323.7 & 5.6910E+08 & 327.9 &  5.0252E+08 & 331.4 &  6.1069E+08 & 325.6 &  2.8894E+08 & 335.4 &  2.0179E+08 &  336.2 \\
2018-04-02 09:27:54 & 8.0212E+08 &  349.6 & 7.4147E+08 & 350.8 &  6.6170E+08 & 352.8 &  7.9098E+08 & 350.4 &  3.8039E+08 & 351.9 &  2.6473E+08 &  350.7 \\
2011-03-14 10:32:05 & 3.6310E+08 &  280.0 & 3.1963E+08 & 286.6 &  2.7418E+08 & 292.1 &  3.5296E+08 & 282.7 &  1.5047E+08 & 301.9 &  1.0074E+08 &  304.1 \\

\noalign{\smallskip}
    \hline
\noalign{\smallskip}


NOAA-19/HIRS4    & \multicolumn{2}{c}{13.347} & \multicolumn{2}{c}{11.124} &  \multicolumn{2}{c}{9.729} & \multicolumn{2}{c}{12.456} &  \multicolumn{2}{c}{7.352} &  \multicolumn{2}{c}{6.529} \\
2017-12-02 12:50:03 & 8.7990E+08 &  359.9 & 8.1692E+08 & 360.0 &  7.3541E+08 & 361.7 &  8.7290E+08 & 360.9 &  4.3089E+08 &  359.0 &  3.0001E+08 &  357.3 \\
2012-03-04 05:07:04 & NaN        &  NaN   &  NaN       &  NaN  &   NaN       &  NaN  &   NaN       &  NaN  &   NaN       &   NaN  &   NaN       &   NaN  \\

\noalign{\smallskip}
    \hline
\noalign{\smallskip}


MetOp-A/HIRS4    & \multicolumn{2}{c}{13.354} & \multicolumn{2}{c}{11.129} &  \multicolumn{2}{c}{9.723} & \multicolumn{2}{c}{12.486} &  \multicolumn{2}{c}{7.343} &  \multicolumn{2}{c}{6.536} \\
2011-11-16 23:42:26 & 3.8384E+08 &  283.9 & 3.4381E+08 & 291.1 &  2.9640E+08 & 296.6 &  3.7100E+08 & 286.6 &  1.6563E+08 & 305.8 &  1.1339E+08 &  308.8 \\

\noalign{\smallskip}
    \hline
\noalign{\smallskip}


MetOp-B/HIRS4    & \multicolumn{2}{c}{13.390} & \multicolumn{2}{c}{11.127} &  \multicolumn{2}{c}{9.715} & \multicolumn{2}{c}{12.438} &  \multicolumn{2}{c}{7.355} &  \multicolumn{2}{c}{6.521} \\
2019-07-21 07:57:26 & 5.9359E+08 &  321.6 &  5.3386E+08 & 324.5 &  4.6917E+08 & 328.1 &  5.8126E+08 & 323.5 &  2.6547E+08 &  331.0 &  1.8268E+08 &  332.1 \\
2015-02-07 05:12:40 & 7.6711E+08 &  344.7 &  7.0041E+08 & 345.6 &  6.2470E+08 & 348.2 &  7.5540E+08 & 345.9 &  3.6176E+08 &  347.8 &  2.5389E+08 &  348.2 \\

      \noalign{\smallskip}
      \hline
    \end{tabular}
}
    \end{center}
  \end{table}
\end{landscape}

\begin{landscape}
  \begin{table}[h!tb]
    \begin{center}
{\tiny
    \caption{Calibrated Radiances and brightness temperatures of the Moon for the short-wavelength channels 13-19. No brightness temperatures
             were determined for channels 17 and 18 due to the unknown contribution of reflected sunlight.
             \label{tbl:hirs_obs_app_tbl3}}
    \begin{tabular}{l|rrrrrrrrrrrr}
      \hline
      \hline
      \noalign{\smallskip}
Satellite/Instrument & \multicolumn{2}{c}{$\lambda_{ch13}$} & \multicolumn{2}{c}{$\lambda_{ch14}$} & \multicolumn{2}{c}{$\lambda_{ch15}$} & \multicolumn{2}{c}{$\lambda_{ch16}$} & \multicolumn{2}{c}{$\lambda_{ch17}$} & \multicolumn{1}{c}{$\lambda_{ch18}$} & \multicolumn{1}{c}{$\lambda_{ch19}$ [$\mu$m]} \\
Observation Epoch    & Rad [MJy/sr] & T$_b$ [K]        & Rad [MJy/sr] & T$_b$ [K]        & Rad [MJy/sr] & T$_b$ [K]        & Rad [MJy/sr] & T$_b$ [K]        & Rad [MJy/sr] & T$_b$ [K]        & Rad [MJy/sr] & Rad [MJy/sr] \\
        
      \noalign{\smallskip}
      \hline
      \noalign{\smallskip}

NOAA-11/HIRS2    &  \multicolumn{2}{c}{4.566} &  \multicolumn{2}{c}{4.525} &  \multicolumn{2}{c}{4.466} &  \multicolumn{2}{c}{4.409} &  \multicolumn{2}{c}{4.138} &  \multicolumn{1}{c}{3.981} &  \multicolumn{1}{c}{3.753} \\
1993-11-25 17:39 & 4.4544E+07 & 344.6 & 4.2184E+07 & 344.6 & 3.9387E+07 & 345.1 & 3.7092E+07 & 345.9 & 2.6219E+07 & 348.7 & 2.1502E+07 & 1.5759E+07 \\
1992-02-15 19:54 & NaN        & NaN   & NaN        & NaN   & NaN        & NaN   & NaN        & NaN   & 2.8494E+07 & 351.7 & NaN        & NaN          \\
1990-11-27 05:04 & NaN        & NaN   & NaN        & NaN   & NaN        & NaN   & NaN        & NaN   & 1.1669E+07 & 322.5 & NaN        & NaN          \\
1989-03-17 03:31 & NaN        & NaN   & NaN        & NaN   & NaN        & NaN   & NaN        & NaN   & 1.6397E+07 & 333.0 & NaN        & NaN          \\
\noalign{\smallskip}
    \hline
\noalign{\smallskip}

NOAA-14/HIRS2    &  \multicolumn{2}{c}{4.563} &  \multicolumn{2}{c}{4.530} &  \multicolumn{2}{c}{4.472} &  \multicolumn{2}{c}{4.409} &  \multicolumn{2}{c}{4.132} &  \multicolumn{1}{c}{3.980} &  \multicolumn{1}{c}{3.776} \\
2000-04-14 10:50:44 & 3.2097E+07 & 332.8 &  3.0767E+07 & 333.0 &  2.8551E+07 & 333.3 &  2.6388E+07 & 333.9 &  1.8025E+07 & 336.4 &   1.4746E+07 &  1.1002E+07 \\
1997-12-08 21:52:12 & 2.2330E+07 & 320.5 &  2.1375E+07 & 320.7 &  1.9854E+07 & 321.2 &  1.8307E+07 & 321.8 &  1.2494E+07 & 324.9 &   1.0216E+07 &  7.6185E+06 \\
1997-06-15 05:05:20 & 1.9721E+07 & 316.5 &  1.9015E+07 & 317.0 &  1.7633E+07 & 317.5 &  1.6281E+07 & 318.2 &  1.1215E+07 & 321.7 &   9.1837E+06 &  6.8734E+06 \\
1996-10-23 00:30:08 & 3.8316E+07 & 339.1 &  3.6903E+07 & 339.4 &  3.4323E+07 & 339.8 &  3.1810E+07 & 340.4 &  2.2115E+07 & 343.2 &   1.8199E+07 &  1.3676E+07 \\
1996-05-28 20:58:58 & 3.0631E+07 & 331.2 &  2.9515E+07 & 331.5 &  2.7481E+07 & 332.0 &  2.5460E+07 & 332.7 &  1.7827E+07 & 336.1 &   1.4694E+07 &  1.1107E+07 \\
1995-12-03 07:13:24 & 4.5724E+07 & 345.7 &  4.3947E+07 & 345.9 &  4.0874E+07 & 346.2 &  3.7817E+07 & 346.6 &  2.6089E+07 & 348.9 &   2.1415E+07 &  1.6047E+07 \\

\noalign{\smallskip}
    \hline
\noalign{\smallskip}

NOAA-15/HIRS3    &  \multicolumn{2}{c}{4.570} &  \multicolumn{2}{c}{4.525} &  \multicolumn{2}{c}{4.474} &  \multicolumn{2}{c}{4.460} &  \multicolumn{2}{c}{4.134} &  \multicolumn{1}{c}{3.970} &  \multicolumn{1}{c}{3.763} \\
2019-12-14 09:47:24 &  NaN       & NaN   &  NaN        & NaN   &  NaN        & NaN   &  NaN        & NaN   &  NaN       & NaN   &   2.7599E+07 &  2.0883E+07 \\
2018-12-24 11:44:46 & 6.4095E+07 & 358.6 &  6.0583E+07 & 358.7 &  5.6851E+07 & 358.8 &  5.6030E+07 & 359.0 & 3.7900E+07 & 362.3 &   3.0009E+07 &  2.2534E+07 \\

\noalign{\smallskip}
    \hline
\noalign{\smallskip}

NOAA-17/HIRS3    &  \multicolumn{2}{c}{4.575} &  \multicolumn{2}{c}{4.523} &  \multicolumn{2}{c}{4.478} &  \multicolumn{2}{c}{4.464} &  \multicolumn{2}{c}{4.138} &  \multicolumn{1}{c}{3.970} &  \multicolumn{1}{c}{3.763} \\
2003-01-24 01:20:02 & 1.9038E+07 & 314.8 &  1.7845E+07 & 315.3 &  1.6969E+07 & 316.0 &  1.6659E+07 & 316.1 &  1.1227E+07 & 321.4 &   8.9445E+06 &  6.7361E+06 \\
2002-09-26 07:01:22 & 3.6261E+07 & 336.5 &  3.3903E+07 & 336.7 &  3.2094E+07 & 337.1 &  3.1419E+07 & 337.1 &  2.0808E+07 & 340.8 &   1.6245E+07 &  1.2074E+07 \\

\noalign{\smallskip}
    \hline
\noalign{\smallskip}

NOAA-18/HIRS4    &  \multicolumn{2}{c}{4.568} &  \multicolumn{2}{c}{4.528} &  \multicolumn{2}{c}{4.468} &  \multicolumn{2}{c}{4.451} &  \multicolumn{2}{c}{4.134} &  \multicolumn{1}{c}{3.975} &  \multicolumn{1}{c}{3.751} \\
2018-11-26 17:59:54 & 4.6002E+07 & 345.7 &   NaN       & NaN   &  NaN         & NaN   & NaN        & NaN   & 2.5536E+07  & 348.0 &  2.0359E+07  & 1.4623E+07 \\
2018-04-02 09:27:54 &  NaN       & NaN   &   NaN       & NaN   &  NaN         & NaN   & NaN        & NaN   & 3.5377E+07  & 359.8 &  2.8333E+07  & 2.0688E+07 \\
2011-03-14 10:32:05 & 2.0236E+07 & 317.1 &  1.9064E+07 & 317.1 &  1.7484E+07  & 317.4 & 1.7137E+07 & 317.6 & 1.1122E+07  & 321.3 &  8.9085E+06  & 6.4512E+06 \\

\noalign{\smallskip}
    \hline
\noalign{\smallskip}


NOAA-19/HIRS4    &    \multicolumn{2}{c}{nan} &    \multicolumn{2}{c}{nan} &    \multicolumn{2}{c}{nan} &    \multicolumn{2}{c}{nan} &  \multicolumn{2}{c}{4.131} &  \multicolumn{1}{c}{3.971} &  \multicolumn{1}{c}{3.757} \\
2017-12-02 12:50:03 &  NaN   & NaN &  NaN   & NaN &   NaN  & NaN &    NaN  & NaN &   3.8682E+07 & 363.3 &   3.1112E+07 &  2.3276E+07 \\
2012-03-04 05:07:04 &  NaN   & NaN &  NaN   & NaN &   NaN  & NaN &    NaN  & NaN &    NaN       & NaN   &   1.3562E+07 &  9.9920E+06 \\

\noalign{\smallskip}
    \hline
\noalign{\smallskip}


MetOp-A/HIRS4    &  \multicolumn{2}{c}{4.567} &  \multicolumn{2}{c}{4.520} &  \multicolumn{2}{c}{4.469} &  \multicolumn{2}{c}{4.453} &  \multicolumn{2}{c}{4.134} &  \multicolumn{1}{c}{3.974} &  \multicolumn{1}{c}{3.754} \\
2011-11-16 23:42:26 & 2.3734E+07 & 322.3 &  2.2566E+07 & 323.0 &  2.0971E+07 & 323.1 &  2.0538E+07 & 323.3 &  1.3263E+07 & 326.6 &   1.0328E+07 &  7.6021E+06 \\
\noalign{\smallskip}
    \hline
\noalign{\smallskip}


MetOp-B/HIRS4    &  \multicolumn{2}{c}{4.575} &  \multicolumn{2}{c}{4.532} &  \multicolumn{2}{c}{4.476} &  \multicolumn{2}{c}{4.458} &  \multicolumn{2}{c}{4.130} &  \multicolumn{1}{c}{3.975} &  \multicolumn{1}{c}{3.753} \\
2019-07-21 07:57:26 & 4.0649E+07 & 340.7 &  3.7884E+07 & 340.3 &  3.5236E+07 & 340.5 &  3.4523E+07 & 340.7 &  2.2187E+07 & 343.5 &   1.7703E+07 &  1.2787E+07 \\
2015-02-07 05:12:40 &  NaN       & NaN   &  5.5484E+07 & 354.8 &  5.1837E+07 & 355.1 &  5.0795E+07 & 355.2 &  3.3225E+07 & 357.7 &   2.6680E+07 &  1.9463E+07 \\
      \noalign{\smallskip}
      \hline
    \end{tabular}
}
    \end{center}
  \end{table}
\end{landscape}


\begin{landscape}
  \begin{table}[h!tb]
    \begin{center}
{\tiny
    \caption{Extracted disk-integrated fluxes of the Moon for the long-wavelength channels 1-12.
             \label{tbl:hirs_obs_app_long}}
    \begin{tabular}{l|rrrrrrrrrrrr}
      \hline
      \hline
      \noalign{\smallskip}
            Satellite/Instrument & $\lambda_{ch01}$ & $\lambda_{ch02}$ & $\lambda_{ch03}$ & $\lambda_{ch04}$ & $\lambda_{ch05}$ & $\lambda_{ch06}$ & $\lambda_{ch07}$ & $\lambda_{ch08}$ & $\lambda_{ch09}$ & $\lambda_{ch10}$ & $\lambda_{ch11}$ & $\lambda_{ch12}$ [$\mu$m] \\
            Observation Epoch    & flx$_1$ & flx$_2$ & flx$_3$ & flx$_4$ & flx$_5$ & flx$_6$ & flx$_7$ & flx$_8$ & flx$_9$ & flx$_{10}$ & flx$_{11}$ & flx$_{12}$ [Jy]\\
        
      \noalign{\smallskip}
      \hline
      \noalign{\smallskip}

NOAA-11/HIRS2    & 14.852 & 14.731 & 14.488 & 14.183 & 13.909 & 13.658 & 13.340 & 11.088 &  9.733 & 12.565 &  7.362 &  6.738 \\
1993-11-25 17:39 & 4.012E+10 & 4.124E+10 & 4.120E+10 & 4.071E+10 & 4.008E+10 & 4.007E+10 & 4.004E+10 & 3.615E+10 & 3.251E+10 & 3.897E+10 & 1.842E+10 & 1.399E+10  \\
1992-02-15 19:54 & 5.388E+10 & 5.565E+10 & 5.584E+10 & 5.505E+10 & 5.433E+10 & 5.462E+10 & 5.439E+10 & 4.964E+10 & 4.472E+10 & 5.326E+10 & 2.540E+10 & 1.929E+10  \\
1990-11-27 05:04 & 2.659E+10 & 2.703E+10 & 2.742E+10 & 2.677E+10 & 2.645E+10 & 2.632E+10 & 2.624E+10 & 2.299E+10 & 2.017E+10 & 2.537E+10 & 1.103E+10 & 8.137E+09  \\
1989-03-17 03:31 & 2.995E+10 & 3.137E+10 & 3.194E+10 & 3.160E+10 & 3.125E+10 & 3.133E+10 & 3.111E+10 & 2.798E+10 & 2.484E+10 & 3.034E+10 & 1.377E+10 & 1.028E+10  \\
\noalign{\smallskip}
    \hline
\noalign{\smallskip}
NOAA-14/HIRS2    & 14.955 & 14.701 & 14.487 & 14.171 & 13.931 & 13.656 & 13.317 & 11.119 &  9.725 & 12.508 &  7.398 &  6.733 \\
2000-04-14 10:50 & 3.663E+10 & 3.698E+10 & 3.708E+10 & 3.704E+10 & 3.725E+10 & 3.712E+10 & 3.685E+10 & 3.256E+10 & 2.808E+10 & 3.566E+10 & 1.648E+10 & 1.230E+10 \\
1997-12-08 21:52 & 2.841E+10 & 2.873E+10 & 2.898E+10 & 2.879E+10 & 2.900E+10 & 2.897E+10 & 2.876E+10 & 2.518E+10 & 2.156E+10 & 2.776E+10 & 1.249E+10 & 9.223E+09 \\
1997-06-15 05:05 & 2.195E+10 & 2.316E+10 & 2.356E+10 & 2.335E+10 & 2.357E+10 & 2.364E+10 & 2.353E+10 & 2.082E+10 & 1.785E+10 & 2.281E+10 & 1.006E+10 & 7.356E+09 \\
1996-10-23 00:30 & 4.380E+10 & 4.343E+10 & 4.351E+10 & 4.356E+10 & 4.376E+10 & 4.374E+10 & 4.315E+10 & 3.873E+10 & 3.395E+10 & 4.189E+10 & 1.976E+10 & 1.484E+10 \\
1996-05-28 20:58 & 3.653E+10 & 3.516E+10 & 3.509E+10 & 3.502E+10 & 3.517E+10 & 3.510E+10 & 3.449E+10 & 3.078E+10 & 2.686E+10 & 3.335E+10 & 1.531E+10 & 1.134E+10 \\
1995-12-03 07:13 & 4.162E+10 & 4.195E+10 & 4.253E+10 & 4.257E+10 & 4.287E+10 & 4.296E+10 & 4.276E+10 & 3.870E+10 & 3.395E+10 & 4.173E+10 & 2.021E+10 & 1.518E+10 \\
\noalign{\smallskip}
    \hline
\noalign{\smallskip}
NOAA-15/HIRS3 & 14.976 & 14.743 & 14.508 & 14.226 & 13.963 & 13.666 & 13.357 & 11.151 &  9.705 & 12.444 &  7.303 &  6.534 \\
2019-12-14 09:47 &       NAN &       NAN &       NAN &       NAN &       NAN &       NAN &       NAN &       NAN &       NAN &       NAN &       NAN &       NAN \\
2018-12-24 11:44 & 5.708E+10 & 5.757E+10 & 5.757E+10 & 5.774E+10 & 5.797E+10 & 5.802E+10 & 5.748E+10 & 5.367E+10 & 4.420E+10 & 5.707E+10 & 2.628E+10 & 1.921E+10 \\
\noalign{\smallskip}
    \hline
\noalign{\smallskip}
NOAA-17/HIRS3 & 14.966 & 14.669 & 14.474 & 14.219 & 13.958 & 13.664 & 13.360 & 11.123 &  9.722 & 12.430 &  7.320 &  6.547 \\
2003-01-24 01:19 & 2.385E+10 & 2.445E+10 & 2.438E+10 & 2.478E+10 & 2.460E+10 & 2.459E+10 & 2.414E+10 & 2.115E+10 & 1.683E+10 & 2.358E+10 & 9.439E+09 & 6.543E+09 \\
2002-09-26 07:01 & 2.853E+10 & 2.976E+10 & 3.036E+10 & 3.090E+10 & 3.109E+10 & 3.151E+10 & 3.148E+10 & 2.932E+10 & 2.443E+10 & 3.124E+10 & 1.385E+10 & 9.738E+09 \\
\noalign{\smallskip}
    \hline
\noalign{\smallskip}
NOAA-18/HIRS4 & 14.980 & 14.697 & 14.513 & 14.223 & 14.006 & 13.670 & 13.340 & 11.117 &  9.720 & 12.501 &  7.314 &  6.513 \\
2018-11-26 17:59 & 4.125E+10 & 4.311E+10 & 4.332E+10 & 4.324E+10 & 4.317E+10 & 4.314E+10 & 4.320E+10 & 3.965E+10 & 3.501E+10 & 4.255E+10 & 2.013E+10 & 1.406E+10 \\
2018-04-02 09:27 & 4.911E+10 & 4.985E+10 & 4.997E+10 & 4.991E+10 & 4.988E+10 & 5.002E+10 & 5.030E+10 & 4.650E+10 & 4.150E+10 & 4.960E+10 & 2.385E+10 & 1.660E+10 \\
2011-03-14 10:32 & 2.413E+10 & 2.399E+10 & 2.399E+10 & 2.385E+10 & 2.371E+10 & 2.369E+10 & 2.370E+10 & 2.086E+10 & 1.789E+10 & 2.304E+10 & 9.820E+09 & 6.575E+09 \\
\noalign{\smallskip}
    \hline
\noalign{\smallskip}

NOAA-19/HIRS4 & 14.953 & 14.685 & 14.526 & 14.232 & 13.973 & 13.635 & 13.347 & 11.124 &  9.729 & 12.456 &  7.352 &  6.529 \\
2017-12-02 12:49 & 6.311E+10 & 6.316E+10 & 6.316E+10 & 6.286E+10 & 6.273E+10 & 6.302E+10 & 6.330E+10 & 5.877E+10 & 5.290E+10 & 6.279E+10 & 3.100E+10 & 2.158E+10 \\
2012-03-04 05:07 &       NAN &       NAN &       NAN &       NAN &       NAN &       NAN &       NAN &       NAN &       NAN &       NAN &       NAN &       NAN \\
\noalign{\smallskip}
    \hline
\noalign{\smallskip}

MetOp-A/HIRS4 & 14.955 & 14.719 & 14.503 & 14.248 & 13.955 & 13.667 & 13.354 & 11.129 &  9.723 & 12.486 &  7.343 &  6.536 \\
2011-11-16 23:42 & 2.407E+10 & 2.428E+10 & 2.439E+10 & 2.435E+10 & 2.442E+10 & 2.431E+10 & 2.421E+10 & 2.168E+10 & 1.869E+10 & 2.340E+10 & 1.045E+10 & 7.151E+09 \\
\noalign{\smallskip}
    \hline
\noalign{\smallskip}

MetOp-B/HIRS4 & 14.964 & 14.688 & 14.474 & 14.248 & 13.996 & 13.671 & 13.390 & 11.127 &  9.715 & 12.438 &  7.355 &  6.521 \\
2019-07-21 07:57 & 3.414E+10 & 3.431E+10 & 3.414E+10 & 3.407E+10 & 3.384E+10 & 3.378E+10 & 3.387E+10 & 3.046E+10 & 2.677E+10 & 3.316E+10 & 1.515E+10 & 1.042E+10 \\
2015-02-07 05:12 & 4.396E+10 & 4.382E+10 & 4.386E+10 & 4.377E+10 & 4.354E+10 & 4.354E+10 & 4.369E+10 & 3.989E+10 & 3.558E+10 & 4.303E+10 & 2.060E+10 & 1.446E+10 \\
      \noalign{\smallskip}
      \hline
    \end{tabular}
}
    \end{center}
  \end{table}
\end{landscape}

\begin{landscape}
  \begin{table}[h!tb]
    \begin{center}
{\tiny
    \caption{Extracted disk-integrated fluxes of the Moon for the short-wavelength channels 13-19.
             \label{tbl:hirs_obs_app_short}}
    \begin{tabular}{l|rrrrrrrrrrrr|rrrrrrr}
      \hline
      \hline
      \noalign{\smallskip}
            Satellite/Instrument & $\lambda_{ch13}$ & $\lambda_{ch14}$ & $\lambda_{ch15}$ & $\lambda_{ch16}$ & $\lambda_{ch17}$ & $\lambda_{ch18}$ & $\lambda_{ch19}$ [$\mu$m] \\
            Observation Epoch    & flx$_{13}$ & flx$_{14}$ & flx$_{15}$ & flx$_{16}$ & flx$_{17}$ & flx$_{18}$ & flx$_{19}$ [Jy]\\
        
      \noalign{\smallskip}
      \hline
      \noalign{\smallskip}

NOAA-11/HIRS2    &  4.566 &  4.525 &  4.466 &  4.409 &  4.138 &  3.981 &  3.753 \\
1993-11-25 17:39 & 2.552E+09 & 2.417E+09 & 2.257E+09 & 2.125E+09 & 1.502E+09 & 1.232E+09 & 9.029E+08 \\
1992-02-15 19:54 &       NAN &       NAN &       NAN &       NAN & 2.053E+09 &       NAN &       NAN \\
1990-11-27 05:04 &       NAN &       NAN &       NAN &       NAN & 7.640E+08 &       NAN &       NAN \\
1989-03-17 03:31 &       NAN &       NAN &       NAN &       NAN & 9.814E+08 &       NAN &       NAN \\
\noalign{\smallskip}
    \hline
\noalign{\smallskip}
NOAA-14/HIRS2    &  4.563 &  4.530 &  4.472 &  4.409 &  4.132 &  3.980 &  3.776 \\
2000-04-14 10:50 & 2.117E+09 & 2.029E+09 & 1.883E+09 & 1.741E+09 & 1.189E+09 & 9.726E+08 & 7.257E+08 \\
1997-12-08 21:52 & 1.535E+09 & 1.470E+09 & 1.365E+09 & 1.259E+09 & 8.590E+08 & 7.023E+08 & 5.238E+08 \\
1997-06-15 05:05 & 1.163E+09 & 1.122E+09 & 1.040E+09 & 9.605E+08 & 6.616E+08 & 5.418E+08 & 4.055E+08 \\
1996-10-23 00:30 & 2.642E+09 & 2.545E+09 & 2.367E+09 & 2.194E+09 & 1.525E+09 & 1.255E+09 & 9.431E+08 \\
1996-05-28 20:58 & 1.949E+09 & 1.878E+09 & 1.749E+09 & 1.620E+09 & 1.134E+09 & 9.350E+08 & 7.067E+08 \\
1995-12-03 07:13 & 2.747E+09 & 2.641E+09 & 2.456E+09 & 2.272E+09 & 1.568E+09 & 1.287E+09 & 9.642E+08 \\
\noalign{\smallskip}
    \hline
\noalign{\smallskip}
NOAA-15/HIRS3 &  4.570 &  4.525 &  4.474 &  4.460 &  4.134 &  3.970 &  3.763 \\
2019-12-14 09:47 &       NAN &       NAN &       NAN &       NAN &       NAN & 1.832E+09 & 1.386E+09 \\
2018-12-24 11:44 & 4.603E+09 & 4.351E+09 & 4.083E+09 & 4.024E+09 & 2.722E+09 & 2.155E+09 & 1.618E+09 \\
\noalign{\smallskip}
    \hline
\noalign{\smallskip}
NOAA-17/HIRS3 &  4.575 &  4.523 &  4.478 &  4.464 &  4.138 &  3.970 &  3.763 \\
2003-01-24 01:19 & 1.305E+09 & 1.223E+09 & 1.163E+09 & 1.142E+09 & 7.696E+08 & 6.131E+08 & 4.617E+08 \\
2002-09-26 07:01 & 2.103E+09 & 1.966E+09 & 1.862E+09 & 1.822E+09 & 1.207E+09 & 9.422E+08 & 7.003E+08 \\
\noalign{\smallskip}
    \hline
\noalign{\smallskip}
NOAA-18/HIRS4 &  4.568 &  4.528 &  4.468 &  4.451 &  4.134 &  3.975 &  3.751 \\
2018-11-26 17:59 & 3.205E+09 &       NAN &       NAN &       NAN & 1.779E+09 & 1.419E+09 & 1.019E+09 \\
2018-04-02 09:27 &       NAN &       NAN &       NAN &       NAN & 2.219E+09 & 1.777E+09 & 1.297E+09 \\
2011-03-14 10:32 & 1.321E+09 & 1.244E+09 & 1.141E+09 & 1.118E+09 & 7.259E+08 & 5.814E+08 & 4.210E+08 \\
\noalign{\smallskip}
    \hline
\noalign{\smallskip}

NOAA-19/HIRS4 &    nan &    nan &    nan &    nan &  4.131 &  3.971 &  3.757 \\
2017-12-02 12:49 &      NAN &       NAN &       NAN &       NAN & 2.783E+09 & 2.238E+09 & 1.674E+09 \\
2012-03-04 05:07 &      NAN &       NAN &       NAN &       NAN &       NAN & 8.511E+08 & 6.270E+08 \\
\noalign{\smallskip}
    \hline
\noalign{\smallskip}

MetOp-A/HIRS4 &  4.567 &  4.520 &  4.469 &  4.453 &  4.134 &  3.974 &  3.754 \\
2011-11-16 23:42 & 1.497E+09 & 1.423E+09 & 1.323E+09 & 1.295E+09 & 8.365E+08 & 6.514E+08 & 4.795E+08 \\
\noalign{\smallskip}
    \hline
\noalign{\smallskip}

MetOp-B/HIRS4 &  4.575 &  4.532 &  4.476 &  4.458 &  4.130 &  3.975 &  3.753 \\
2019-07-21 07:57 & 2.319E+09 & 2.161E+09 & 2.010E+09 & 1.970E+09 & 1.266E+09 & 1.010E+09 & 7.295E+08 \\
2015-02-07 05:12 &       NAN & 3.160E+09 & 2.953E+09 & 2.893E+09 & 1.892E+09 & 1.520E+09 & 1.109E+09 \\
      \noalign{\smallskip}
      \hline
    \end{tabular}
}
    \end{center}
  \end{table}
\end{landscape}

\end{appendix}

\end{document}